\let\mypdfximage\pdfximage
\def\pdfximage{\immediate\mypdfximage}

\documentclass[letterpaper, 11pt]{article}

\usepackage[linesnumbered,ruled]{algorithm2e}
\usepackage{amsfonts}
\usepackage{amsmath}
\usepackage{amssymb}
\usepackage{amsthm}
\usepackage{epsfig}
\usepackage{geometry}
\usepackage{graphicx}
\usepackage[colorlinks=true, citecolor=blue, filecolor=black, linkcolor=red, urlcolor=blue]{hyperref}
\usepackage{subfigure}
\usepackage[textsize=tiny]{todonotes}

\usepackage[capitalise,nameinlink,noabbrev]{cleveref} 

\usepackage{pgfplots}
\usepackage{tikz}

\usetikzlibrary{calc, shapes, arrows, positioning, angles, quotes,
  patterns, fit}

\pgfplotsset{compat=newest} 
\pgfplotsset{plot coordinates/math parser=false} 
\newlength\figureheight 
\newlength\figurewidth

\renewcommand{\[}{\left[}
\renewcommand{\]}{\right]}
\renewcommand{\(}{\left(}
\renewcommand{\)}{\right)}

\newcommand{\norm}[2]{\left\|\, #1 \,\right\|_{#2}}

\newcommand{\vvvert}{|\kern-1pt|\kern-1pt|}

\newcommand{\hf}{\hat{f}}

\newcommand{\talpha}{\tilde{\alpha}}

\newcommand{\EE}{\mathbb{E}}

\newcommand{\NN}{\mathbb{N}}

\newcommand{\CJ}{\mathcal{J}}

\newcommand{\CN}{\mathcal{N}}

\newcommand{\CU}{\mathcal{U}}

\newcommand{\Var}{\textrm{Var}}

\newcommand{\etal}{\textit{et al.}}

\newcommand{\argmin}{\operatornamewithlimits{arg\,min}}

\newcommand{\iid}{\stackrel{\textrm{iid}}{\sim}}

\def\argmin{\mathop{\mathrm{argmin}}}

\crefname{section}{Sec.}{Sec.}
\Crefname{section}{Section}{Sections}
\crefname{subsection}{Sec.}{Sec.}
\Crefname{subsection}{Section}{Sections}
\crefname{figure}{Fig.}{Fig.}
\Crefname{figure}{Figure}{Figures}
\crefname{equation}{Eqn.}{Eqn.}
\Crefname{equation}{Equation}{Equations}

\crefdefaultlabelformat{#2\textup{#1}#3}

\title{Global Sensitivity Analysis and Estimation of Model Error,
  Toward Uncertainty Quantification in Scramjet Computations}

\author{Xun Huan\footnote{Corresponding author:
    \href{mailto:xhuan@sandia.gov}{xhuan@sandia.gov}, Sandia National
    Laboratories, Livermore, CA 94550, USA.}, Cosmin
  Safta\footnote{Sandia National Laboratories, Livermore, CA 94550,
    USA.}, Khachik Sargsyan\footnotemark[2],\\Gianluca
  Geraci\footnote{Sandia National Laboratories, Albuquerque, NM 87123,
    USA.}, Michael S. Eldred\footnotemark[3], Zachary
  P. Vane\footnotemark[2],\\Guilhem Lacaze\footnotemark[2], Joseph
  C. Oefelein\footnotemark[2], and Habib N. Najm\footnotemark[2]}

\begin{document}

\maketitle

\begin{abstract}
  The development of scramjet engines is an important research area
  for advancing hypersonic and orbital flights. Progress toward
  optimal engine designs requires accurate flow simulations together
  with uncertainty quantification. However, performing uncertainty
  quantification for scramjet simulations is challenging due to the
  large number of uncertain parameters involved and the high
  computational cost of flow simulations. These difficulties are
  addressed in this paper by developing practical uncertainty
  quantification algorithms and computational methods, and deploying
  them in the current study to large-eddy simulations of a jet in
  crossflow inside a simplified HIFiRE Direct Connect Rig scramjet
  combustor. First, global sensitivity analysis is conducted to
  identify influential uncertain input parameters, which can help
  reduce the system’s stochastic dimension. Second, because models of
  different fidelity are used in the overall uncertainty
  quantification assessment, a framework for quantifying and
  propagating the uncertainty due to model error is presented. These
  methods are demonstrated on a nonreacting jet-in-crossflow test
  problem in a simplified scramjet geometry, with parameter space up
  to 24 dimensions, using static and dynamic treatments of the
  turbulence subgrid model, and with two-dimensional and
  three-dimensional geometries.
\end{abstract}

\section*{Nomenclature}

\noindent\begin{tabular}{@{}lcl@{}}
$C_R$ &=& modified Smagorinsky constant \\
$c_{[\sim k]}$ &=& solution to the reduced least absolute shrinkage
and selection operator problem \\
$c_{\beta^n}$ &=& coefficient for the $n$th basis function\\
$D$ &=& data set \\
$d$ &=& injector diameter, mm \\
$E_{\mathrm{CV}}$ &=& cross-validation error \\
$f(s, \lambda)$ &=& low-fidelity model \\
$f(\lambda)$ &=& quantity of interest (model output) \\
$f_{\ell}(\lambda)$ &=& quantity of interest from a model with
discretization level $\ell$ \\
$f_{\Delta_{\ell}}(\lambda)$ &=& $f_{{\ell}}(\lambda) -
f_{{\ell-1}}(\lambda)$ \\
$\hf_k(\cdot)$ &=& surrogate model for $f_k(\cdot)$ \\
$\hf_{\ell}, \hf_{\Delta_{\ell}}$ &=& approximations to $f_{\ell}$ and
$f_{\Delta_{\ell}}$ \\
$G, c, f$ &=& regression matrix, solution vector, and right-hand-side vector  \\
$G_{[k]}, f_{[k]}$ &=& $G$ and $f$ with rows corresponding
to the $k$th subset only \\
$G_{[\sim k]}, f_{[\sim k]}$ &=& $G$ and $f$ with rows corresponding to the $k$th subset
removed
\end{tabular}

\noindent\begin{tabular}{@{}lcl@{}}
$g(s)$ &=& high-fidelity model \\
$I_i, I_f$ &=& inlet and fuel turbulence intensity magnitudes \\
$\CJ$ &=& index set  \\
$L_G$ &=& Gaussian approximation to the marginalized likelihood  \\
$L_i, L_f$ &=& inlet and fuel turbulence length scales, mm \\
$\ell$ &=& discretization level \\
$M_0, M_f$ &=& inlet and fuel Mach numbers \\
$\dot{m}_f$ &=& fuel mass flux, kg/s \\
$\iid\CN(\mu,\sigma^2)$ &=& independent and identically distributed as
normal distribution with mean $\mu$\\
&& and variance $\sigma^2$  \\
$n_s$ &=& stochastic dimension  \\
$P_{rms}$ &=& root-mean-square static pressure \\
$P_{stag}$ &=& stagnation pressure \\
$Pr_t$ &=& turbulent Prandtl number \\
$p$ &=& polynomial degree \\
$p(\cdot)$ &=& probability density function \\
$p_0$ &=& inlet stagnation pressure, MPa\\
$q_i$ &=& stochastic model \\
$R_{[k]}$ &=& validation residual to the reduced least absolute
shrinkage and selection \\ & & operator problem \\
$Sc_t$ &=& turbulent Schmidt number \\
$S_i$ &=& main effect Sobol sensitivity index for the $i$th input parameter  \\
$S_{ij}$ &=& joint effect Sobol sensitivity index for the interaction
of $i$th and $j$th input\\&& parameters  \\
$S_{T_i}$ &=& total effect Sobol sensitivity index for all terms
involving the $i$th input parameter  \\
$s$ &=& shared continuous operating conditions \\
$T_{f}$ &=& fuel static temperature, K \\
$T_w$ &=& wall temperature  \\
$T_0$ &=& inlet stagnation temperature, K \\
$\sim\CU(a,b)$ &=& distributed as uniform distribution from $a$ to $b$  \\
$x, y, z$ &=& streamwise, wall-normal, and spanwise coordinates, mm \\
$x/d, y/d, z/d$ &=& streamwise, wall-normal, and spanwise coordinates
normalized by injector diameter \\
$Y_{\textnormal{C}_2\textnormal{H}_4}$  &=& ethylene mass fraction \\
$Z$ &=& mixture fraction \\
$\alpha$ &=& parameter of $\delta$ \\
$\talpha$ &=& combined parameters $\lambda$ and $\alpha$ \\
$\beta$ &=& multi-index \\
$\delta_a$ &=& inlet boundary layer thickness, mm \\
$\delta_i$ &=& model discrepancy term \\
$\epsilon_k$ &=& surrogate model error \\
$\eta$ &=& least absolute shrinkage and selection operator regularization parameter \\
$\lambda$ &=& model input parameter vector \\
$\lambda_i, \lambda_{\sim i}$ &=& $i$th component of $\lambda$, all
components of $\lambda$ except the $i$th \\
$\lambda_{\beta}$ &=& expansion coefficients \\
$\mu$ &=& mean  \\
$\Xi$ &=& support of $p(\xi)$  \\
$\xi_j$ &=& independent and identically distributed basic (germ) random variables \\
$\xi^{(m)}$ &=& $m$th regression point
\end{tabular}

\noindent\begin{tabular}{@{}lcl@{}}
$\rho_i$ &=& multiplicative term\\
$\sigma,\sigma^2$ &=& standard deviation, variance \\
$\chi$ &=& scalar dissipation rate \\
$\Psi_{\beta}, \psi_{\beta_j}$ &=& multivariate and univariate
orthonormal polynomial basis functions \\
$\Psi_{\beta^n}$ &=& $n$th basis function
\end{tabular}

\section{Introduction}
\label{s:intro}

Supersonic combustion ramjet (scramjet) engines allow propulsion
systems to transition from supersonic to hypersonic flight conditions
while ensuring stable combustion, potentially offering much higher
efficiencies compared to traditional technologies such as rockets or
turbojets. While several scramjet designs have been conceived, none to
date operate optimally~\cite{Schmisseur2015}. This is due to
difficulties in characterizing and predicting combustion properties
under extreme flow conditions, coupled with multiscale and
multiphysics nature of the processes involved.  Designing an optimal
engine involves maximizing combustion efficiency while minimizing
pressure losses, thermal loading, and the risk of ``unstart'' or flame
blowout. Achieving this, especially in the presence of uncertainty, is
an extremely challenging undertaking.

An important step towards optimal scramjet design is to conduct
accurate flow simulations together with uncertainty quantification
(UQ). While UQ in general has received substantial attention in the
past decades, UQ for scramjet applications is largely undeveloped,
with a few exceptions~\cite{Witteveen2011, Constantine2015}.  A
comprehensive UQ study in such systems has been prohibitive due to
both the large number of uncertainty sources in the predictive models
as well as the high computational cost of simulating multidimensional
turbulent reacting flows.  This study aims to advance practical
algorithms and computational methods that enable tractable UQ analysis
of realistic scramjet designs. The immediate goals are to:
\begin{enumerate}
\item focus on an initial nonreacting jet-in-crossflow test
  problem in a simplified scramjet geometry;

\item identify influential uncertain input parameters via global
  sensitivity analysis, which can help reduce the system's stochastic
  dimension;

\item quantify and propagate the uncertainty due to model error from
  using low-fidelity models; and

\item demonstrate these UQ methods on the jet-in-crossflow problem
  (nonreacting, simplified scramjet geometry), and prepare extensions
  to its full configuration.
\end{enumerate}

\begin{figure}[hbt]
  \centering
  \mbox{\subfigure[HIFiRE Flight 2 payload]{
      \includegraphics[width=0.6\textwidth]{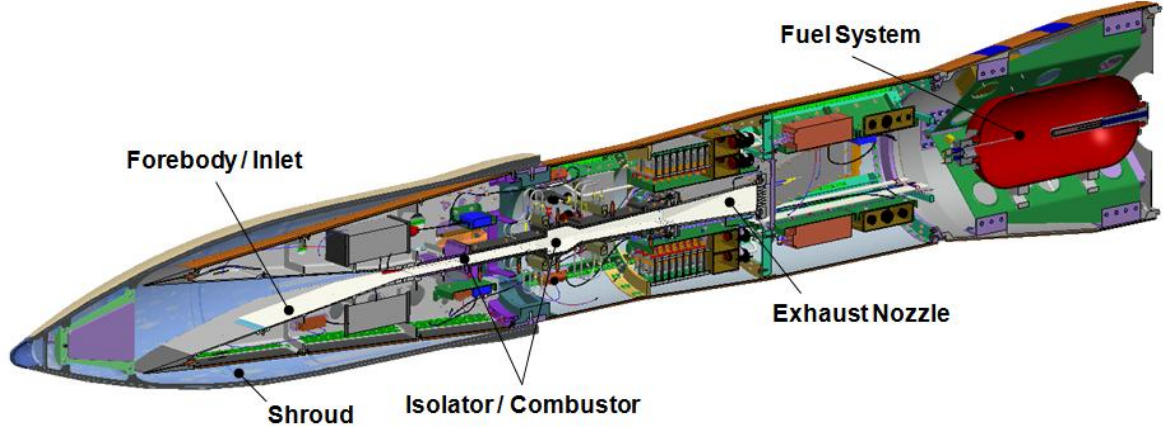}\label{f:LES_SCRAMJET}}
  }
  \mbox{\subfigure[HDCR]{
      \includegraphics[width=0.35\textwidth]{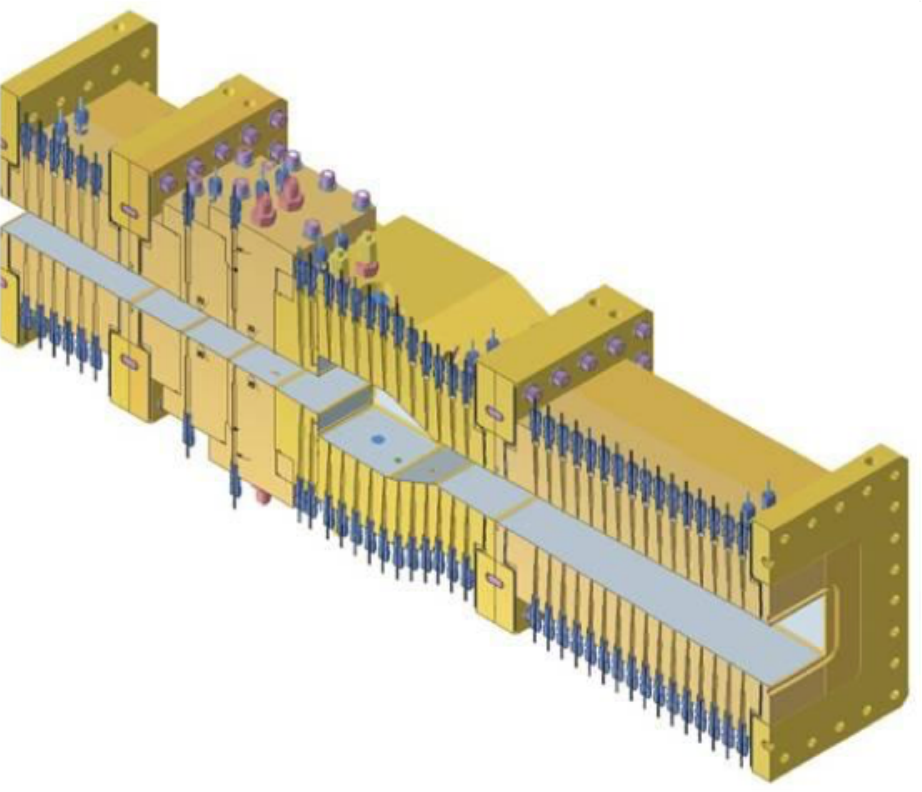}\label{f:LES_experimental}}
  }
        \caption{HIFiRE Flight 2 payload~\cite{Jackson2009} and
          HDCR~\cite{Hass2010} cut views.}
  \label{f:LES}
\end{figure}

We concentrate on a scramjet configuration studied under the HIFiRE
(Hypersonic International Flight Research and Experimentation)
program~\cite{Dolvin2008, Dolvin2009}, which has been the target of a
mature experimental campaign with accessible data through its HIFiRE
Flight 2 (HF2) project~\cite{Jackson2009, Jackson2011}. The HF2
payload, depicted in \cref{f:LES_SCRAMJET}, involves a cavity-based
hydrocarbon-fueled dual-mode scramjet that enables transition from
ramjet mode (subsonic flow in the combustor) to scramjet mode
(supersonic flow in the combustor) through a variable Mach number
flight trajectory. The isolator/combustor was derived from a series of
legacy configurations at the U.S. Air Force Research
Laboratory~\cite{Gruber2008, Jackson2011}, whereas the forebody,
inlet, and nozzle were designed at NASA Langley Research
Center~\cite{Ferlemann2008, Gruber2009}.  A ground test rig,
designated the HIFiRE Direct Connect Rig (HDCR,
\cref{f:LES_experimental}), was developed to duplicate the
isolator/combustor layout of the flight test hardware and to provide
ground-based data for comparisons with flight data, verifying engine
performance and operability as well as designing fuel delivery
schedule~\cite{Hass2010, Storch2011}. Mirroring the HDCR setup, we aim
to simulate and assess flow characteristics inside the
isolator/combustor portion of the scramjet.

The paper is structured as follows. \Cref{s:LES} describes the physics
and solver used for simulating the jet in crossflow inside the
simplified HDCR scramjet combustor, in particular with the use of
large-eddy simulation techniques.  We then introduce global
sensitivity analysis in \cref{s:GSA} to identify the most influential
input parameters of the model. In \cref{s:merr}, a framework is
presented to capture uncertainty from model error when low-fidelity
models are used.  These UQ methods are then demonstrated on the
jet-in-crossflow problem, with results shown in
\cref{s:results}. Finally, the paper ends with conclusions and future
work discussions in \cref{s:conclusions}.

\section{Large-Eddy Simulations for the HIFiRE Direct Connect Rig}
\label{s:LES}

\subsection{Computational domain}

\begin{figure}[htb]
  \centering
  \mbox{\subfigure[HDCR geometry and computational domain]{
        \includegraphics[width=0.45\textwidth]{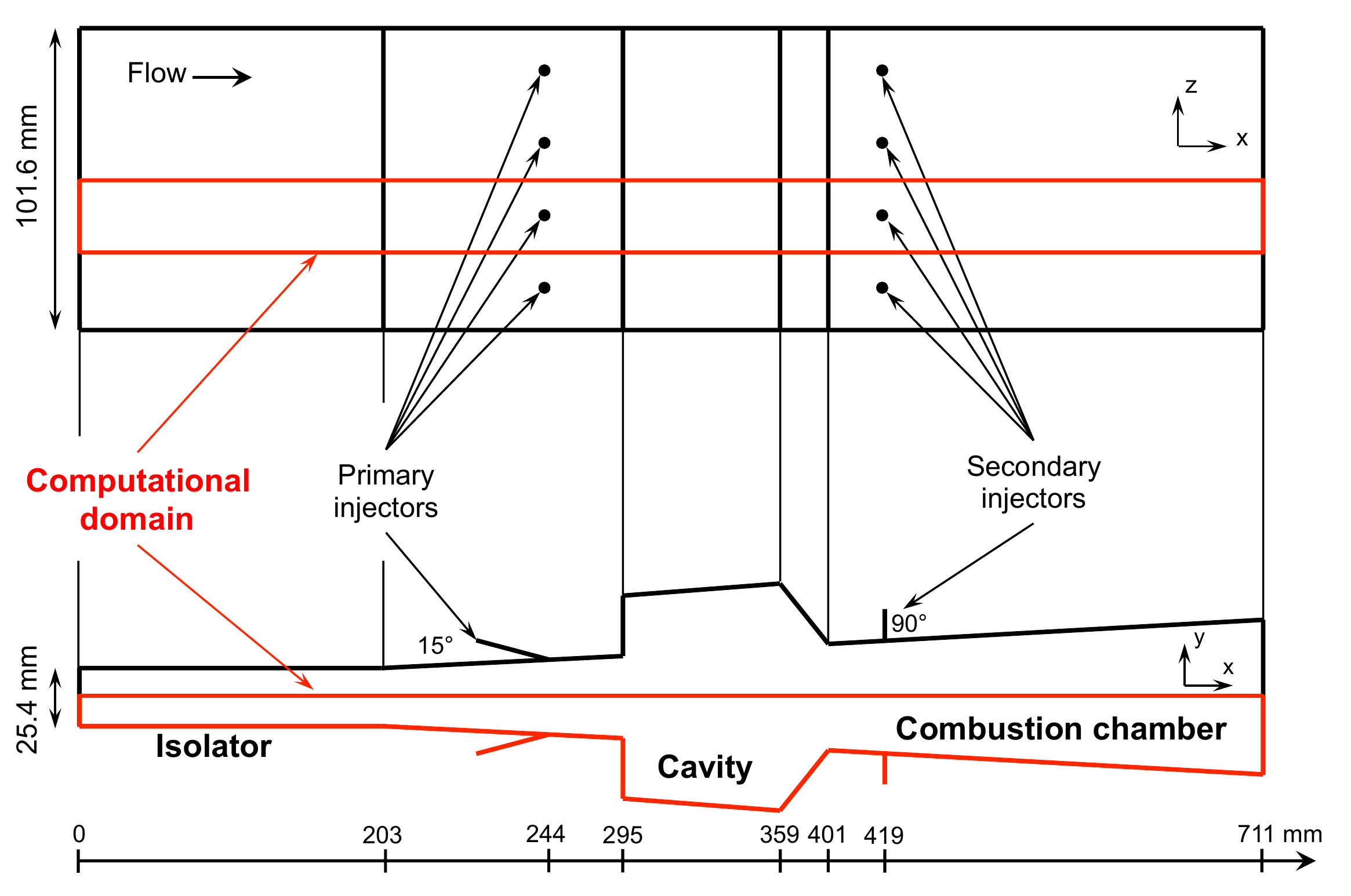}\label{f:LES_full_schematic}}
  }
  \mbox{\subfigure[Unit test problem computational domain]{
      \includegraphics[width=0.5\textwidth]{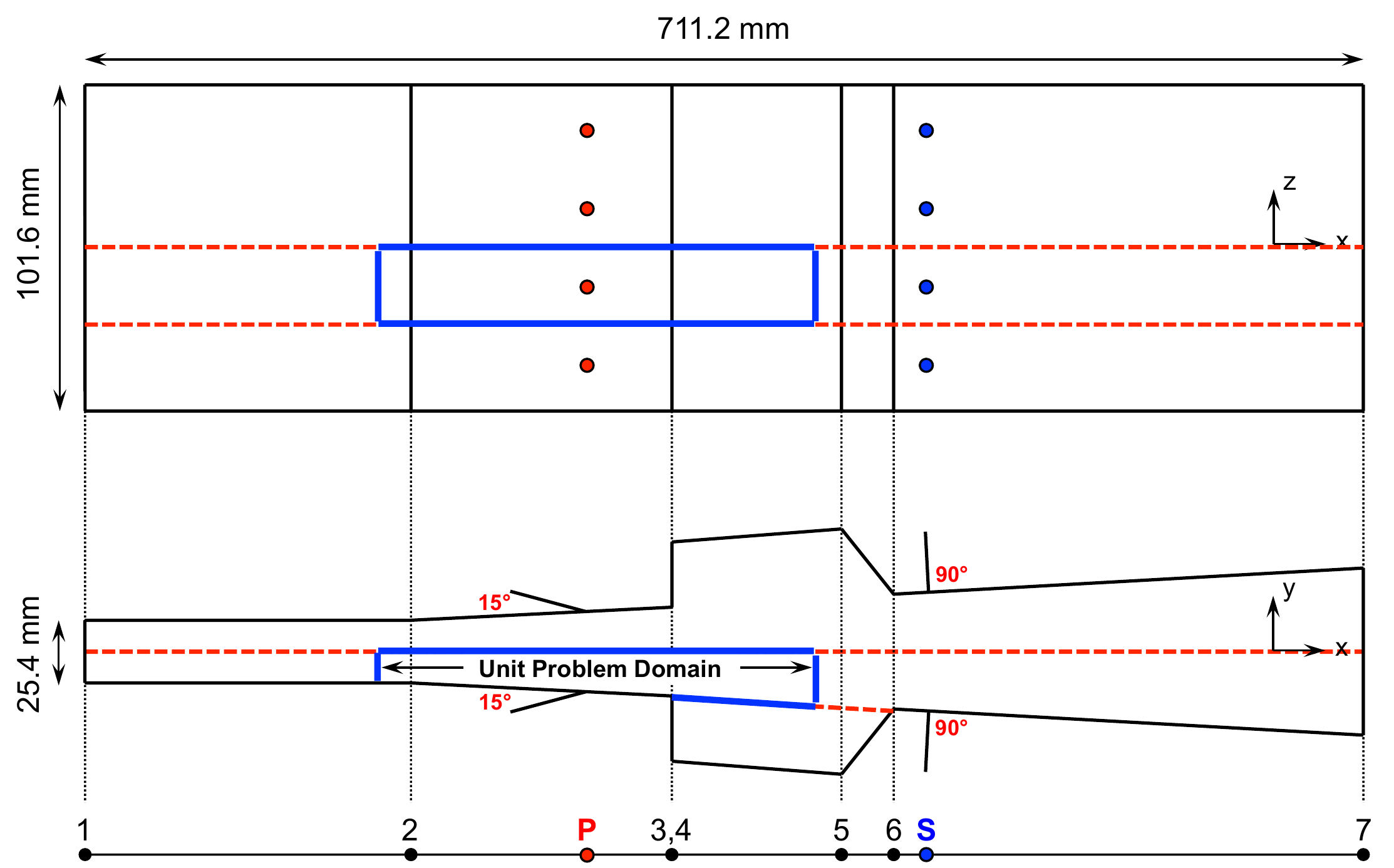}\label{f:LES_p1_schematic}}
  }
  \caption{Schematics of HDCR geometry and computational domain
    (left), and unit test problem computational domain (right,
    highlighted by solid blue lines). }
  \label{f:LES_schematics}
\end{figure}

We aim to perform flow simulations inside the HDCR.  A detailed
schematic of the HDCR geometry is shown in
\cref{f:LES_full_schematic}. The rig consists of a constant-area
isolator (i.e., planar duct) attached to a combustion chamber.  It
includes four primary injectors that are mounted upstream of flame
stabilization cavities on both the top and bottom walls. Four
secondary injectors along both walls are also positioned downstream of
the cavities.  Flow travels from left to right in the $x$-direction
(streamwise), and the geometry is symmetric about the centerlines in
both the $y$-direction (wall-normal) and $z$-direction (spanwise).
Numerical simulations can take advantage of this symmetry by
considering a domain that comprises only the bottom half and one set
of the primary and secondary injectors. This ``full'' computational
domain is highlighted by red lines in \cref{f:LES_full_schematic}.

Even with symmetry-based size reductions of the computational domain,
the cost associated with the thousands of simulations required for UQ
analysis necessitates further simplifications.  Since the current
focal point is to develop, validate, and demonstrate the various UQ
methodologies, a unit test problem is designed for these
purposes. Calculations are performed in the region near the primary
injectors along the bottom wall ($x = 190$ to 350 mm). The domain is
simplified by considering only a single primary injector and omitting
the presence of the cavity. Chemistry is initially disabled, allowing
a targeted investigation of the interaction between the fuel jet (JP-7
surrogate: 36\% methane and 64\% ethylene) and the supersonic
crossflow without the effects of combustion reaction.  The location of
the outflow boundary is chosen to ensure the flow to be fully
supersonic across the entire exit plane. The computational domain for
the unit test problem is identified by the solid blue lines in
\cref{f:LES_p1_schematic}. The flow conditions of interest correspond
to the freestream and fuel injection parameters reported by the HDCR
experiments. Details related to large-eddy simulation of the full HDCR
configuration are described by Lacaze~\etal~\cite{lacaze-aiaa-asm17}.

\subsection{Large-eddy simulation solver: RAPTOR}

Large-eddy simulation (LES) calculations are performed using the
RAPTOR code framework developed by
Oefelein~\cite{oefelein-pas06,oefelein-phd}. The solver has been
optimized to meet the strict algorithmic requirements imposed by the
LES formalism. The theoretical framework solves the fully coupled
conservation equations of mass, momentum, total-energy, and species
for a chemically reacting flow. It is designed to handle high Reynolds
number, high-pressure, real-gas and/or liquid conditions over a wide
Mach operating range. It also accounts for detailed thermodynamics and
transport processes at the molecular level. Noteworthy is that RAPTOR
is designed specifically for LES using non-dissipative, discretely
conservative, staggered, finite-volume differencing. This eliminates
numerical contamination of the subfilter models due to artificial
dissipation and provides discrete conservation of mass, momentum,
energy, and species, which is imperative for high quality LES.
Representative results and case studies using RAPTOR can be found in
studies by Oefelein~\etal~\cite{Oefelein2006a, Oefelein2007,
  Oefelein2012, Oefelein2013, Oefelein2014}, Williams~\etal~\cite{
  Williams2007}, Lacaze~\etal~\cite{Lacaze2015}, and
Khalil~\etal~\cite{Khalil2015}.

\begin{figure}
\centering
\includegraphics[width=0.47\textwidth]{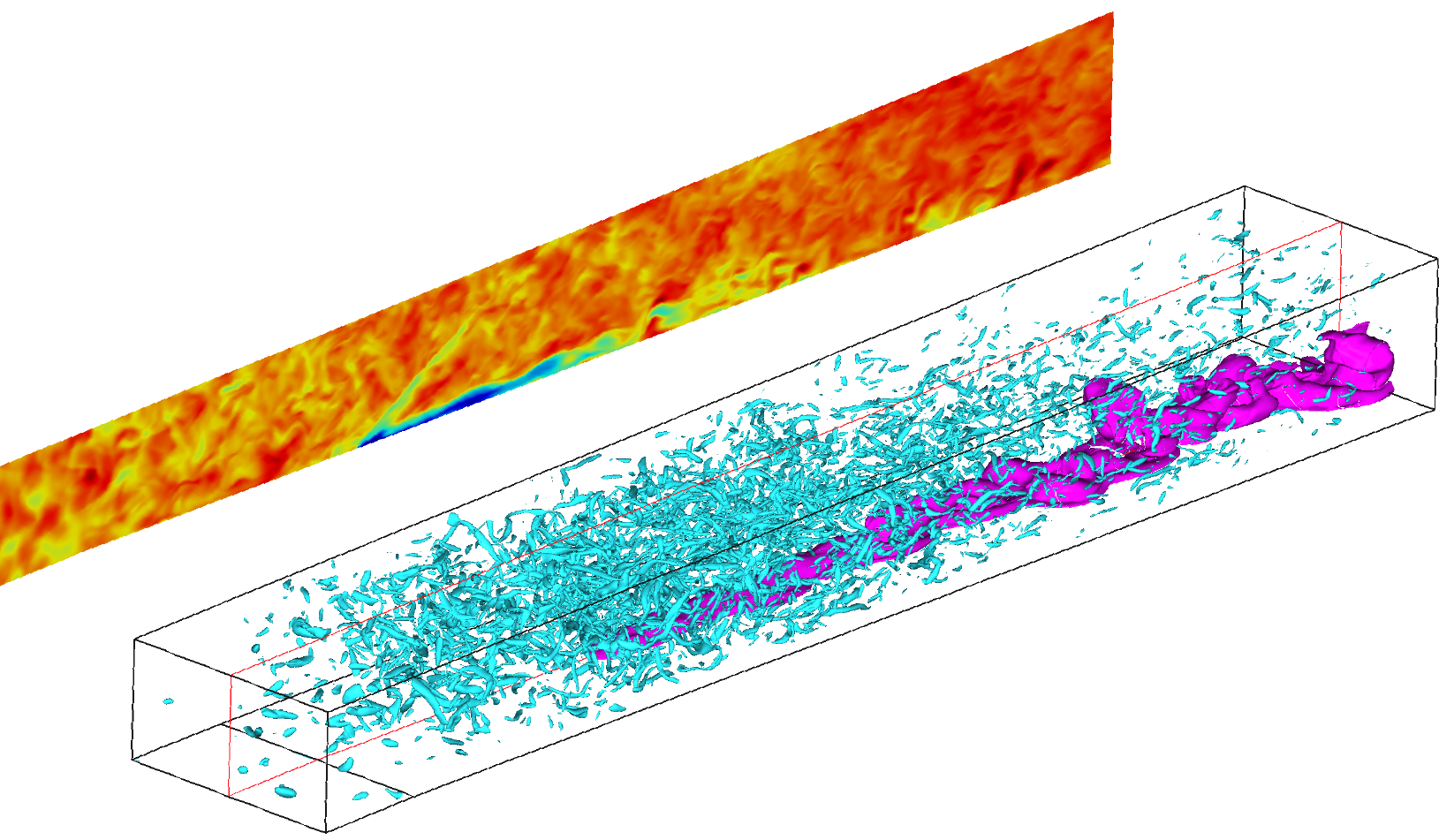}
\includegraphics[width=0.47\textwidth]{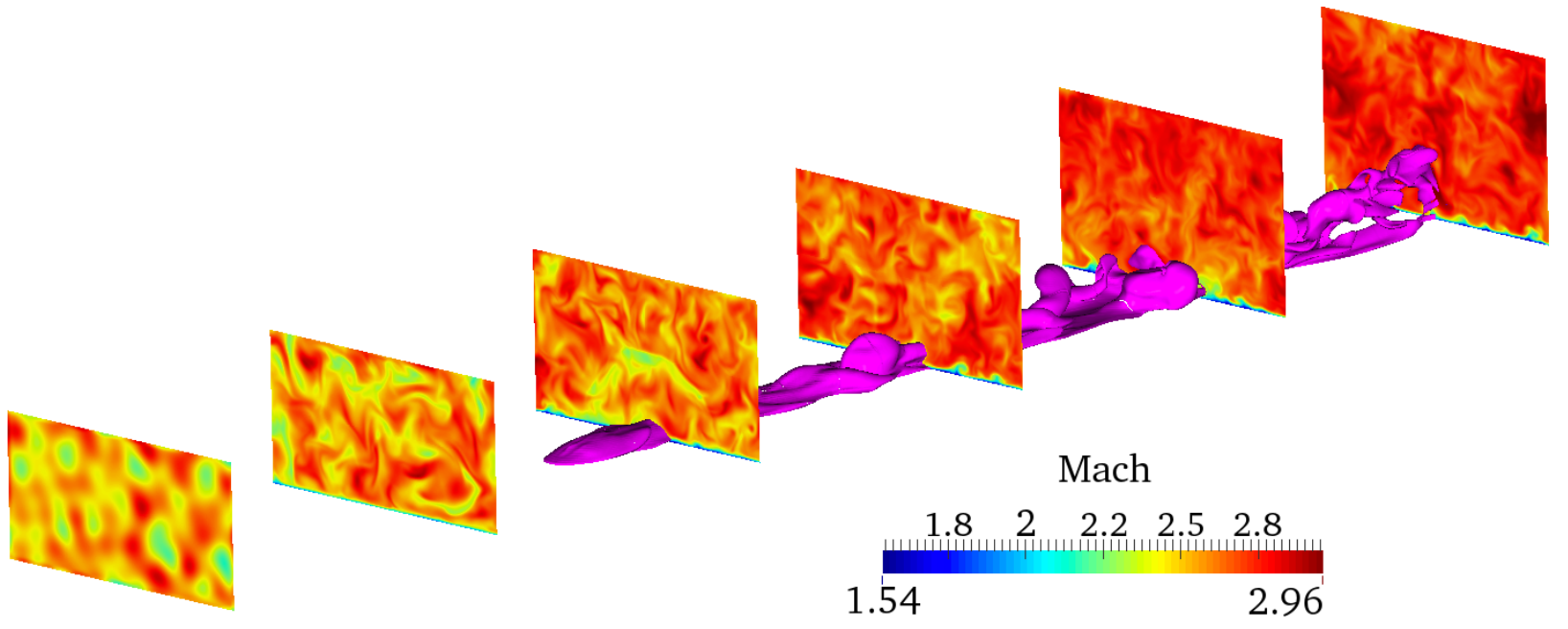}
\caption{Sample results for the jet-in-crossflow test
  problem. Ethylene (fuel component): purple isocontour
  ($Y_{\textnormal{C}_2\textnormal{H}_4} = 0.1$), turbulence: blue
  iso-contour ($Q$-criterion = $10^5~s^{-2}$), and cutting planes are
  colored by the Mach number. }
\label{f:LES_p1}
\end{figure}

Sample results for the jet-in-crossflow problem are presented in
\cref{f:LES_p1}. Here, the instantaneous Mach number is shown across
different planes along with isocontours of ethylene (fuel component)
mass fraction and $Q$-criterion contours that highlight coherent
turbulent structures.

\section{Global Sensitivity Analysis}
\label{s:GSA}

UQ encompasses many different investigations (e.g., uncertainty
propagation, optimal experimental design, model calibration,
optimization under uncertainty); we start by introducing global
sensitivity analysis (GSA)~\cite{Saltelli2004, Saltelli2008} in this
paper.  GSA provides insights on the behavior of uncertainty of model
output quantities of interest (QOIs), and identifies input parameters
that are unimportant to these QOIs, which may be subsequently
eliminated---i.e., it is useful for dimension reduction of the input
space.  GSA achieves this by quantifying the importance of each
uncertain input parameter with respect to the predictive uncertainty
of a given QOI. In contrast to local sensitivity analysis, GSA
reflects the overall sensitivity characteristics across the
\textit{entire} input domain.

We focus on variance-based properties of the input and output
variables. Loosely speaking, variance of a QOI can be decomposed into
contributions from the uncertainty of each input parameter. Let
$\lambda$ denote the vector of all input parameters; we compute Sobol
sensitivity indices~\cite{Sobol2003} to rank the components
$\lambda_i$ in terms of their variance contributions to a given QOI
$f(\lambda)$:
\begin{itemize}
\item \textit{Main effect sensitivity} measures variance contribution
  solely due to the $i$th parameter:
  \begin{align}
    S_i = \frac{\Var_{\lambda_i}\(
      \EE_{\lambda_{\sim i}}\[f(\lambda)|\lambda_i\]\)}{\Var\(f(\lambda)\)}.
  \end{align}
  The notation $\lambda_{\sim i}$ refers to all components of
  $\lambda$ \textit{except} the $i$th component.

\item \textit{Joint effect sensitivity} measures variance contribution
  from the interaction of $i$th and $j$th parameters:
  \begin{align}
    S_{ij} = \frac{\Var_{\lambda_{ij}}\( \EE_{\lambda_{\sim
          ij}}\[f(\lambda)|\lambda_{ij}\]\)}{\Var\(f(\lambda)\)} - S_i
    - S_j.
  \end{align}
      
\item \textit{Total effect sensitivity} measures variance
  contributions from \textit{all} terms that involve the $i$th
  parameter:
  \begin{align}
    S_{T_i} = \frac{\EE_{\lambda_{\sim i}}\[
      \Var_{\lambda_{i}}\(f(\lambda)|\lambda_i\)\]}{\Var\(f(\lambda)\)}.\label{e:sobol_tot}
  \end{align}

\end{itemize}
Total effect sensitivity is particularly informative for identifying
parameters that have the highest overall impact on the QOI. The
unimportant parameters, for example, then may be fixed at their
nominal values without significantly underrepresenting the QOI
variance. Subsequently, the stochastic input dimension would be
reduced at a cost of only small variance approximation
errors. \textit{The primary objective of the GSA study in this paper
  is to compute the total effect sensitivity indices.}

Traditionally, sensitivity indices are directly estimated via various
flavors of efficient Monte Carlo (MC) methods~\cite{Sobol1990,
  Jansen1999, Saltelli1999, Saltelli2002, Saltelli2010}.  The number
of samples needed, however, is typically impractical when expensive
models (such as LES) are involved. We tackle this difficulty via two
approaches. First, we take advantage of multilevel (ML) and
multifidelity (MF) formulations, to construct QOI approximations by
combining simulation from models of different discretization levels
(e.g., grid resolutions) and fidelity (i.e., modeling
assumptions). These frameworks help transfer some of the computational
burden from expensive models to inexpensive ones, and reduce the
overall sampling cost. In particular, a control-variate-based MLMF
Monte Carlo (MLMF MC) method is used to produce efficient sample
allocation across different models.  Second, we adopt polynomial chaos
expansions (PCEs) to approximate the QOIs in ML and MF forms, thereby
presuming a certain degree of smoothness in the QOIs, with attendant
computational savings for given accuracy requirements. Additionally,
compressed sensing (CS) is used to discover sparse PCE structures from
small numbers of model evaluations. Once these PCEs become available,
their orthogonal polynomial basis functions allow Sobol indices to be
extracted analytically from expansion coefficients without the need of
additional MC sampling.

\begin{figure}
  \centering
  \resizebox{0.95\textwidth}{!}{

        \tikzset{      dimen/.style={<->,>=latex,very thick,every rectangle node/.style={midway,font=\footnotesize}},
    }
    
    \tikzstyle{block} = [rectangle, draw, fill=blue!20, 
      text width=20em, text centered, rounded corners, minimum height=4em]
    \tikzstyle{cloud} = [draw, ellipse, text centered, fill=red!20, node distance=3cm,
      minimum height=2em]

    \begin{tikzpicture}[node distance = 7.em, auto]

      \node [block, text width=25em] (GSA) {\textbf{Global sensitivity
          analysis (GSA):} \cite{Saltelli2004, 
            Saltelli2008, Sobol2003}\\\vspace{0.5em}{\small
          \textit{Goal:} Compute Sobol sensitivity indices
          (e.g., total sensitivity indices \cref{e:sobol_tot})}};
      \node [cloud, below of=GSA, text width=6.5em, node distance=10em] (MLMFPCE) {PCEs of ML/MF forms};
      \node [block, left of=MLMFPCE, text width=14em, node
        distance=15em] (MLMF)
            {\textbf{Multilevel/multifidelity\\(ML/MF):} \cite{Giles2008,
                Ng2012, Eldred2015, 
            Peherstorfer2016, Geraci2015,
            Geraci2017}\\\vspace{0.5em}{\small Establish ML/MF forms
          (\cref{e:ML_telescopic}), by
          combining simulations from multiple models
          of different resolution levels and fidelity while trading off
          between computational cost and accuracy}};
      \node [block, right of=MLMFPCE, text width=16em, node
        distance=15em] (PCE) {\textbf{Polynomial chaos\\ 
          expansions (PCEs):} \cite{Ghanem1991, Najm2009, Xiu2009,
            LeMaitre2010, Ernst2012, Xiu2002}\\\vspace{0.5em}{\small
          Compute PCE coefficients through regression
          (\cref{e:PCE_regression}); then extract Sobol indices 
          analytically from coefficients (e.g., total sensitivity
          indices from \cref{e:STi_from_PCE})}};
      \node [block, below of=PCE, node distance=11em] (CS) {\textbf{Compressed sensing
          (CS):} \cite{Candes2006a,
          Donoho2006a}\\\vspace{0.5em}{\small Discover sparse PCE
          structures from small
        numbers of model evaluations, by solving the LASSO
        problem (\cref{e:Lagrangian_LASSO}) (e.g., using
        GPSR~\cite{Figueiredo2007}) and choosing regularization parameter
      that minimizes CV error (\cref{e:CV_opt})}};

            \draw [dimen,<-] (GSA) -- (MLMFPCE);
      \draw [dimen,<-] (MLMFPCE) -- (MLMF);
      \draw [dimen,<-] (MLMFPCE) -- (PCE);
      \draw [dimen,<-] (PCE) -- (CS);

    \end{tikzpicture}
  }
  \caption{Summary of the computational concepts and tools used for
    GSA.}
  \label{f:GSA_flow}
\end{figure}
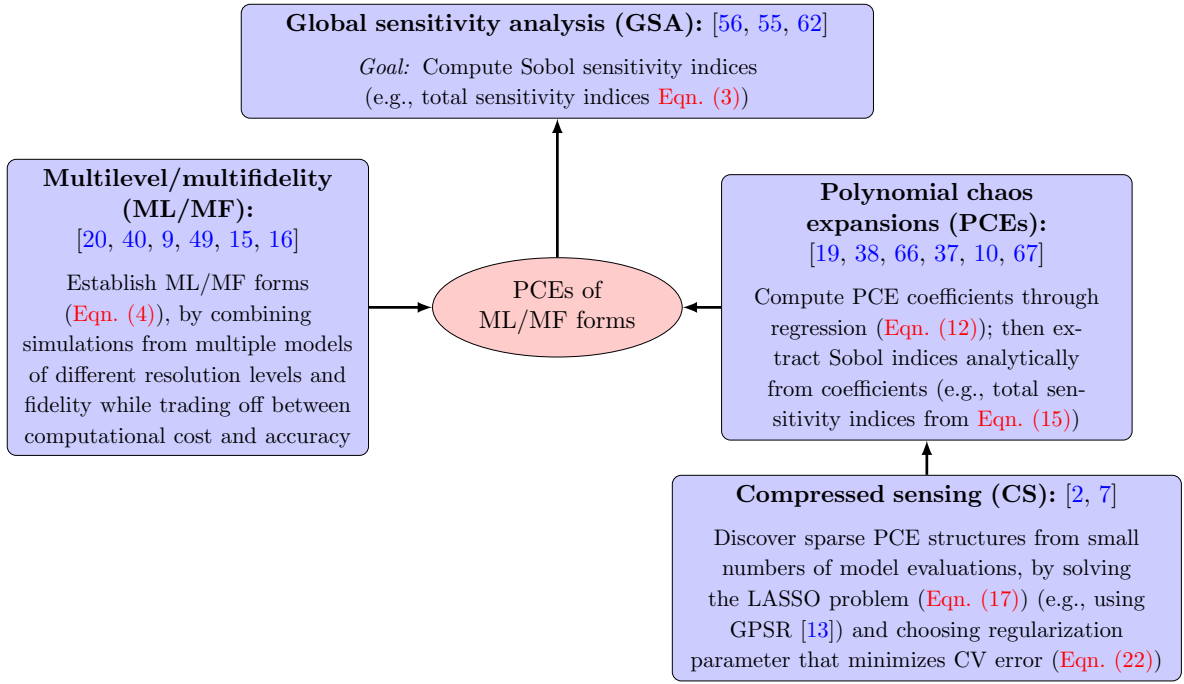

\Cref{f:GSA_flow} shows a summary of the computational concepts and
tools used for GSA, and may serve as a useful reference for readers as
they are introduced later.

\subsection{Multilevel and multifidelity representations}
\label{ss:ml_mf}

Performing UQ directly on the highest fidelity model available is
usually very challenging, if not altogether intractable, due to its
high simulation cost.  Often, models employing different
discretization levels and modeling assumptions are also available, and
even those with coarse grids and low fidelity can provide some
information about the output behavior.  It would be useful to then
extract information from inexpensive simulations whenever possible,
and resort to expensive ones for details that can only be
characterized through those models.  We thus seek multimodel
approaches that are more efficient overall, and specifically focus on
ML and MF representations.

We start by first describing the ML concept.  Consider a generic QOI
produced by a model with discretization level (e.g., grid resolution)
$\ell$, denoted by $f_{\ell}(\lambda)$, and $\ell=0$ and $\ell=L$ are
the coarsest and finest available resolutions, respectively. The QOI
from the finest resolution can be expanded exactly as
\begin{align}
  f_{L}(\lambda) = f_{0}(\lambda) + \sum_{\ell = 1}^L
  f_{\Delta_{\ell}}(\lambda),\label{e:ML_telescopic}
\end{align}
where $f_{\Delta_{\ell}}(\lambda) \equiv f_{{\ell}}(\lambda) -
f_{{\ell-1}}(\lambda)$ denotes the difference terms on adjacent
levels. One motivation behind using this telescopic decomposition is
that $f_{\Delta_{\ell}}(\lambda)$ can become better behaved and easier
to characterize than the QOI directly as some of the nonlinear
behavior may be subtracted out.  One can also view this as an
exploitation of the correlation between evaluations at different
levels, where information injected from the inexpensive simulations
help reduce the need to evaluate expensive models, and thus decrease
the overall cost of characterizing $f_L(\lambda)$.

From here, the expansion can be utilized in a few different ways; we
focus on its use for \textit{uncertainty propagation}. For example,
one may take expectations with respect to $\lambda$ on each term, and
generate MC samples to obtain efficient moment estimators of the
finest-resolution QOI. This is known as the ML Monte Carlo method and
has extensive theoretical developments stemming from the work of
Giles~\cite{Giles2008}. Alternatively, one may be interested in
producing functional approximations of the QOI response:
\begin{align}
  f_{L}(\lambda) \approx \hat{f}_{L}(\lambda) = \hat{f}_{0}(\lambda) +
  \sum_{\ell = 1}^L
  \hat{f}_{\Delta_\ell}(\lambda),\label{e:ML_PCE_telescopic}
\end{align}
where $\hat{f}_{\ell}$ and $\hat{f}_{\Delta_\ell}$ are approximations
to $f_{\ell}$ and $f_{\Delta_\ell}$, respectively. We take this path,
and adopt PCEs (see next section) for these approximations. One
motivation for this choice is that we want to leverage the observed
smoothness of QOIs over the parameter space. Another reason is that
the resulting PCEs provide a convenient form in which GSA may be
performed, and the PCEs can also be reused in other UQ investigations
within the overall project.

An analogous argument can be made across models of different fidelity,
and we refer to this parallel expansion as the MF form. The analysis
of MF approaches can be more difficult than ML, since differences
induced by modeling assumptions are more challenging to characterize
systematically, and no longer rely on properties resulting from the
convergence of grid resolution. Nonetheless, MF remains a valuable
tool, and has been initially explored with the use of sparse
grids~\cite{Ng2012, Eldred2015}. A comprehensive survey regarding MF
methods can be found in a report by
Peherstorfer~\etal~\cite{Peherstorfer2016}.

With details in the subsequent section, we proceed to use a
sample-based regression approach to construct the approximations
$\hat{f}_0(\lambda)$ and $\hat{f}_{\Delta_{\ell}}(\lambda)$.  The
allocation of samples evaluated at different levels and fidelity are
computed using the MLMF MC method~\cite{Geraci2015, Geraci2017}. This
algorithm applies a MF control variate to an ML expansion, and
generates a sample allocation that minimizes the variance of the
overall MC estimator, which, at the same time, also accounts for the
computational cost of model simulations. While this allocation
procedure provides an optimal-variance MC estimator, it is not
directly aimed for an optimal construction of the approximation
functions. However, they still provide a good general sample
allocation that is useful for our study here.  Once
$\hat{f}_0(\lambda)$ and $\hat{f}_{\Delta_{\ell}}(\lambda)$ become
available, the overall approximation $\hat{f}_L(\lambda)$ can be
recovered by adding them according to \cref{e:ML_PCE_telescopic}.

\subsection{Polynomial chaos expansion}
\label{ss:pce}

PCEs are used to approximate the terms in \cref{e:ML_PCE_telescopic}.
A PCE is a spectral representation of a random variable. It provides a
useful means for propagating uncertainty as an alternative to MC
simulations. We provide a brief description of PCE below, and refer
readers to several books and review papers for detailed
discussions~\cite{Ghanem1991, Najm2009, Xiu2009, LeMaitre2010}.

With mild technical assumptions~\cite{Ernst2012}, a real-valued random
variable $\lambda$ with finite variance (such as an uncertain input
parameter) can be expanded in the following form:
\begin{align}
  \lambda = \sum_{\norm{\beta}{1}=0}^{\infty} \lambda_{\beta}
  \Psi_{\beta}(\xi_1,\ldots,\xi_{n_s}),
  \label{e:PCEForm2}
\end{align}
where $\xi_j$ are independent and identically distributed (i.i.d.)
basic (germ) random variables; $n_s$ is the stochastic dimension
(often chosen to equal the system stochastic degrees of freedom for
convenience, though \cref{e:PCEForm2} can hold for any finite or
infinite $n_s$ with associated requirements~\cite{Ernst2012});
$\lambda_{\beta}$ are the expansion coefficients;
$\beta=\(\beta_1,\ldots,\beta_{n_s}\),\,\forall \beta_j\in\NN_0$, is a
multi-index; and $\Psi_{\beta}$ are multivariate normalized orthogonal
polynomials written as products of univariate orthonormal polynomials
\begin{align}
  \Psi_{\beta}(\xi_1,\ldots,\xi_{n_s}) = \prod_{j=1}^{n_s}
  \psi_{\beta_j}(\xi_j).
\end{align}
The univariate functions $\psi_{\beta_j}$ are polynomials of degree
$\beta_j$ in the independent variable $\xi_j$, and orthonormal with
respect to the density of $\xi$ (i.e., $p\(\xi\)$):
\begin{align}
  \EE\[\psi_k(\xi)\psi_n(\xi)\] = \int_{\Xi}
  \psi_k\(\xi\)\psi_n\(\xi\)p\(\xi\)\,d\xi =
  \delta_{k,n},
  \label{e:orthogonality}
\end{align}
where $\Xi$ is the support of $p\(\xi\)$. Different choices of $\xi$
and $\psi_{\beta}$ under the generalized Askey family are
available~\cite{Xiu2002}. We employ uniform $\xi\sim \CU(-1,1)$ and
Legendre polynomials in this study.  In practice, the infinite sum in
the expansion \cref{e:PCEForm2} is truncated:
\begin{align}
  \lambda \approx \sum_{\beta \in \CJ} \lambda_{\beta}
  \Psi_{\beta}(\xi_1,\ldots,\xi_{n_s}),
  \label{e:PCEForm3}
\end{align}
where $\CJ$ is some finite index set. For example, one popular choice
for $\CJ$ is the ``total-order'' expansion of degree $p$, where
$\CJ=\{\beta : \norm{\beta}{1} \leq p \}$.
Similarly, we can write the PCE for a QOI in the form
\begin{align}
  f \approx \sum_{\beta \in \CJ} c_{\beta}
  \Psi_{\beta}(\xi_1,\ldots,\xi_{n_s}).
  \label{e:PCEForm4}
\end{align}
Methods for computing its coefficients are broadly divided into two
groups: intrusive and non-intrusive. The former involves substituting
the expansions directly into the governing equations and applying
Galerkin projection, resulting in a larger, new system for the PCE
coefficients that needs to be solved only once. The latter involves
finding an approximation in the subspace spanned by the basis
functions, which typically requires evaluating the original model many
times. With our model only available as a black box in practice, and
also to accommodate flexible choices of QOIs that may be complicated
functions of the state variables, we elect to take the non-intrusive
route.

One non-intrusive method relies on Galerkin projection of the
solution, known as the non-intrusive spectral projection (NISP)
method:
\begin{align}
  c_{\beta} = \EE\[f(\lambda)
    \Psi_{\beta}\] = \int_{\Xi}
    f\(\lambda(\xi)\) \Psi_{\beta}(\xi) p(\xi)
    \,d\xi.
    \label{e:NISP}
\end{align}
Generally, the integral must be estimated numerically and
approximately via, for example, sparse
quadrature~\cite{Barthelmann2000, Gerstner1998, Gerstner2003}. When
the dimension of $\xi$ is high, model is expensive, and only few
evaluations are available, however, even sparse quadrature becomes
impractical. In these situations, regression is a more effective
method, which involves solving the following regression linear system
$Gc = f$:
\begin{align}
  \underbrace{\begin{bmatrix} \Psi_{\beta^1}(\xi^{(1)}) & \cdots &
  \Psi_{\beta^N}(\xi^{(1)}) \\ \vdots & & \vdots \\
  \Psi_{\beta^1}(\xi^{(M)}) & \cdots & 
  \Psi_{\beta^N}(\xi^{(M)}) \end{bmatrix}}_{G}
  \underbrace{\begin{bmatrix} c_{\beta^1} \\  \vdots
  \\ c_{\beta^N} \end{bmatrix}}_{c} =
  \underbrace{\begin{bmatrix}{c} f(\lambda(\xi^{(1)})) \\  \vdots
      \\ f(\lambda(\xi^{(M)})) \end{bmatrix}}_{f},
  \label{e:PCE_regression}
\end{align}
where $\Psi_{\beta^n}$ refers to the $n$th basis function,
$c_{\beta^n}$ is the coefficient corresponding to that term, and
$\xi^{(m)}$ is the $m$th regression (training) point. $G$ is thus the
regression matrix where each column corresponds to a basis term and
each row corresponds to a regression point.

For LES, the affordable number of simulations $M$ is expected to be
drastically smaller than the number of basis terms $N$, leading to an
extremely underdetermined system.  For example, a total-order
expansion of degree 3 in 24 dimensions contains $\frac{(3+24)!}{3!24!}
= 2925$ terms, while 2925 of 3D LES with a moderate $d/16$ grid
resolution (i.e., each grid cell is $1/16$ the size of the injector
diameter $d=3.175$ mm) would take more than 64 million CPU hours!
While ML and MF formulations help reduce the number of expensive model
simulations, we also utilize CS (see next section) to discover sparse
structure in the PCE and remove basis terms with low magnitude
coefficients. Once the final PCE for the QOI is established, we can
extract the Sobol indices via the formulae:
\begin{align}
  S_i &= \frac{1}{\Var\(f(\lambda)\)} \sum_{\beta \in \CJ_{S_i}}
  c_{\beta}^2,\, \textrm{where}\,\,\CJ_{i} = \left\{\beta  \in \CJ:
  \beta_i > 0, \beta_k = 0, k \neq i \right\} \\
  S_{{ij}} &= \frac{1}{\Var\(f(\lambda)\)} \sum_{\beta \in \CJ_{J_{ij}}}
  c_{\beta}^2,\, \textrm{where}\,\, \CJ_{{ij}} = \left\{\beta \in \CJ: \beta_i > 0,
  \beta_j > 0, \beta_k = 0, k \neq i, k \neq j \right\} \\
    S_{T_i} &= \frac{1}{\Var\(f(\lambda)\)} \sum_{\beta \in \CJ_{T_i}}
  c_{\beta}^2,\, \textrm{where} \,\, \CJ_{T_i} = \left\{\beta  \in \CJ:
  \beta_i > 0 \right\}, \label{e:STi_from_PCE}
\end{align}
and the QOI variance can be computed by
\begin{align}
  \Var\(f(\lambda)\) = \sum_{0\neq \beta \in \CJ} c_{\beta}^2.\label{e:PCE_variance}
\end{align}

\subsection{Compressed sensing}

CS~\cite{Candes2006a, Donoho2006a} aims to recover sparse solutions of
underdetermined linear systems, and its use for finding sparse PCEs in
the presence of limited data has received considerable attention
within the UQ community in recent years~\cite{Rauhut2012, Hampton2015,
  Fajraoui2017TR, Peng2014, Eldred2015, Sargsyan2014, Jakeman2015,
  Huan2017c}. Typically, this entails finding the solution with the
fewest number of non-zero components---i.e., minimizing its
$\ell_0$-norm.  However, $\ell_0$-minimization is an NP-hard
problem~\cite{Natarajan1995}. A simpler convex relaxation minimizing
the $\ell_1$-norm minimization is often used as an approximation, and
is proven to uniquely achieve the $\ell_0$ solution in the limit of
large systems and when the true solution is sufficiently
sparse~\cite{Donoho2006}.  We focus on one variant of $\ell_1$-sparse
recovery---the (unconstrained) least absolute shrinkage and selection
operator (LASSO) problem:
\begin{align}
  \min_c \frac{1}{2} \norm{Gc - f}{2}^2 + \eta
  \norm{c}{1}, \label{e:Lagrangian_LASSO}
\end{align}
where $\eta\geq 0$ is a scalar regularization parameter.
We demonstrate one possible method for solving the LASSO problem
(other algorithms were explored as well but omitted for brevity)
through the gradient projection for sparse reconstruction
(GPSR)~\cite{Figueiredo2007}.  GPSR targets \cref{e:Lagrangian_LASSO}
by employing a positive-negative split of the solution vector,
yielding a quadratic program in the resulting new form. A gradient
descent with backtracking is then performed, and constraints are
handled by projection onto the feasible space. For our numerical
demonstrations, we use the MATLAB implementation GPSR v6.0 from the
developers' website~\cite{Figueiredo2009}.

By promoting sparsity, CS is designed to reduce overfitting.  An
overfit solution is observed when the error on training set (i.e.,
data used to define the underdetermined linear system) is very
different (much smaller) than error on a separate validation set, and
the use of a different training set could lead to entirely different
results. Such a solution has poor predictive capability and thus
unreliable. CS is not always successful in preventing overfitting,
such as when $\eta$ in \cref{e:Lagrangian_LASSO} is poorly chosen.
$\eta$ reflects the relative importance between the $\ell_1$ and
$\ell_2$ terms; the former represents regularization and smoothing,
and the latter for producing predictions that closely match the
training data. A large $\eta$ heavily penalizes nonzero terms of the
solution vector, forcing them toward zero (underfitting); a small
$\eta$ emphasizes data fit, and may lead to solutions that are not
sparse, and that \textit{only} fit the training points but otherwise
do not predict well (overfitting).  A useful solution thus requires an
intricate selection of $\eta$, which is a problem-dependent and
nontrivial task. We examine and control the degree of overfitting by
employing cross-validation (CV)~\cite{Hastie2009} to guide the choice
of $\eta$. In particular, we use the $K$-fold CV error.  The procedure
involves first partitioning the full set of $M$ training points into
$K$ (approximately) equal subsets.  For each of the subsets, a reduced
version of the original LASSO problem is solved:
\begin{align}
  c_{[\sim k]}(\eta) = \argmin_{c} \frac{1}{2}\norm{G_{[\sim k]}c -
    f_{[\sim k]}}{2}^2 + \eta
  \norm{c}{1}, \label{e:Lagrangian_LASSO_reduced}
\end{align}
where $G_{[\sim k]}$ denotes $G$ but with rows corresponding the $k$th
subset removed, $f_{[\sim k]}$ is $f$ with elements corresponding to
the $k$th subset removed, and $c_{[\sim k]}(\eta)$ is the solution
vector from solving this reduced problem. The $\ell_2$ residual from
validation using the $k$th subset that was left out is therefore
\begin{align}
  R_{[k]}(\eta) \equiv \norm{G_{[k]} c_{[\sim k]}(\eta)-f_{[k]}}{2},
\end{align}
where $G_{[k]}$ denotes $G$ that only contains rows corresponding to
the $k$th subset, and $f_{[k]}$ is $f$ containing only elements
corresponding to the $k$th subset. Combining the residuals from
repeating the exercise on all $K$ subsets, we arrive at the
(normalized) $K$-fold CV error
\begin{align}
 E_{\textnormal{CV}}(\eta) \equiv
 \frac{\sqrt{\sum_{k=1}^{K}\[R_{[k]}(\eta)\]^2}}{\norm{f}{2}}.\label{e:CV_kfold}
\end{align}
The CV error thus provides an estimate of the validation error using
only the training data set at hand and without needing additional
validation points, and reflects the solution predictive capability.
The CS problem with $\eta$ selection through CV error is:
\begin{align}
  &\min_c \frac{1}{2} \norm{Gc - f}{2}^2 + \eta^{\ast}
  \norm{c}{1},\\
  &\textnormal{where}\hspace{0.5em}\eta^{\ast} = \argmin_{\eta \geq 0}
  E_{\textnormal{CV}}(\eta).\label{e:CV_opt}
\end{align}
Note that solving \cref{e:CV_opt} does not require the solution from
the full linear system, only the $K$ reduced systems.
$\eta^{\ast}$ can be found by, for example, a grid search across the
$\eta$-space.

\section{Embedded Representation of Model Error}
\label{s:merr}

Classical Bayesian model calibration typically assumes that data are
consistently generated from the model---that is, the model is
correct. In reality, all models are approximations to the truth, and
different models trade off between accuracy and computational cost
with their assumptions and parameterizations. For example,
computational studies of turbulent combustion may employ different
geometry details, flow characteristics, turbulence modeling, grid
resolutions, and even the inclusion or removal of entire physical
dimensions. As we make use of different models, it is crucial not only
to acknowledge and understand---but also to develop the capability to
represent, quantify, attribute, and propagate---the uncertainty due to
model error. We address model error in this study by focusing on the
quantities we ultimately care about in an engineering context: model
predictions.

Consider two models: a high-fidelity model $g(s)$ and a low-fidelity
model $f(s,\lambda)$. Both models are functions of shared continuous
operating conditions $s$, which in the context of the current LES
  studies
consist of the spatial coordinates.
The low-fidelity model also carries parameters
$\lambda$, which may be calibrated to provide requisite ``tuning'' of
the model and potentially providing some compensation for its lower
fidelity.
Furthermore, component notations $g_i$ and $f_i$ denote the $i$th
model observable (i.e., the categorical model output
\textit{variable}, such as temperature, pressure, etc.).
We are interested in the uncertainty incurred in predictive QOIs when
$f(s,\lambda)$ is used in place of $g(s)$.  This entails first the
calibration of the low-fidelity model $f(s,\lambda)$ using data from
the high-fidelity model $g(s)$.

We approach the calibration problem taking a Bayesian perspective.
Kennedy and O'Hagan~\cite{Kennedy2001} pioneered a systematic Bayesian
framework for characterizing model error, by employing a linear
stochastic model $q_i(s) = \rho_i f_i(s,\lambda) + \delta_i(s)$, with
the calibration data $g_i(s)$ interpreted as sample realizations of
$q_i(s)$.  This form uses an additive Gaussian process model
discrepancy term $\delta_i(s)$, and
$\rho_i$ is a constant factor.
While this approach is quite flexible in terms of attaining good fits
for each calibration quantity, it presents difficulties when certain
physical properties are desired in the predictions. First, the
multiplicative and additive structures generally cannot guarantee the
predictive quantity to maintain satisfaction of the underlying
governing equations and physical laws instituted in $f$.
Second, the discrepancy term, $\delta_i(s)$, is not transferable for
prediction of variables outside those used for calibration (i.e., for
predicting $g_l(s)$ with $l\neq i$).

Alternative approaches have been introduced in recent years that
target the aforementioned difficulties. For example,
Sargsyan~\etal~\cite{Sargsyan2015} take a non-intrusive approach and
embed a stochastic variable that represents model discrepancy in the
low-fidelity model parameters, while Oliver~\etal~\cite{Oliver2015}
elect to include a correction term on less-reliable embedded models
that exist within a highly-reliable set of governing equations.
We adopt the approach of Sargsyan~\etal~\cite{Sargsyan2015}, first
embedding the discrepancy term in the model parameters
such that $q_i(s) = f_i(s, \lambda + \delta_i(s))$, then employing a
parametric stochastic representation to arrive at the following
characterization of the high-fidelity data behavior:
\begin{align}
  q_i(s) = f_i(s, \lambda + \delta_i(s, \alpha_i, \xi_i)).
  \label{e:embed_general}
\end{align}
Here, $\alpha_i$ is a parameter of $\delta_i(\cdot)$, and $\xi_i$ is a
fixed random variable such as a standard normal.  The uncertainty due
to model error is thus encoded in both the distributions of $\alpha_i$
and $\xi_i$, where the former is reducible and can be learned from
data while the latter remains fixed. In this form, the predictive
quantity automatically preserves physical laws imposed in $f$ to the
extent that this random perturbation of $\lambda$ is within physical
bounds.  Furthermore, while $\delta_i(\cdot)$ remains specific to the
$i$th observable, we expect it to be better behaved, and generally
more meaningful when extrapolated to other observables. This is
supported by the fact that $\delta_i(\cdot)$ is now a correction term
to the same parameter $\lambda$ regardless of $i$, and
$\delta_i(\cdot)$ would always remain in the same physical unit and
likely similar magnitude. In contrast, the additive $\delta_i(s)$ from
Kennedy and O'Hagan would be under entirely different physical units
and potentially orders-of-magnitude different across different
$i$.\footnote{Another known difficulty with the external additive
  model error is the non-idenfiability between model error and
  measurement noise contributions: there can be (possibly infinite)
  different combinations of model discrepancy and data noise that
  together characterize the overall data distribution. The embedded
  form in \cref{e:embed_general} would be able to better separate and
  distinguish the model error (internal) and data noise (external). As
  discussed in \cref{f:time_avg_GSA,f:time_avg_merr} from
  \cref{s:results}, one source of data noise in this paper's numerical
  examples is the variation due to time averaging. However, this
  variation is observed to be very small compared to uncertainty
  contributions from model error and parameter posterior. We thus do
  not include them in the equations and numerical results presented,
  and also do not demonstrate the separation of data noise in the
  embedded representation.}  Nonetheless, extrapolation of $\delta_i$
is still needed for prediction.  Finding the relationship of model
error across different observables is a challenging task. We take a
reasonable first step and use a constant extrapolation, by assuming
$\delta_i$ to be the same across space $s$ (thus it is now a random
variable rather than a random process), and also across all
observables $i$, leading to
\begin{align}
  q_i(s) = f_i(s, \lambda + \delta(\alpha,\xi)).
  \label{e:embed_extrapolate}
\end{align}
Finally, for simplifying notation, $s$ is assumed to be discretized
with nodes $s_j$, and the overall model output vector thus has the
form
\begin{align}
  q_k = f_k(\lambda + \delta(\alpha,\xi)),
  \label{e:embed_discretize}
\end{align}
where $k$ is the combined index of $i$ and $j$.

In this study, we choose to represent the randomly perturbed parameter
as a Legendre-Uniform PCE:
\begin{align}
  \lambda + \delta(\alpha,\xi) = \lambda + \sum_{\beta \neq 0}
  \alpha_{\beta} \Psi_{\beta}(\xi),\label{e:lambda_PCE}
\end{align}
where $\Psi_{\beta}(\xi)$ are Legendre polynomial functions. We use
uniform distributions to better control the range of the perturbed
parameter, though other PCE variants (e.g., Gauss-Hermite) are
possible as well. When $\lambda$ is multi-dimensional, the different
components of $\delta(\cdot)$ may use different orders of
expansions. In practice, we may choose to embed in certain targeted
parameter components, while keeping others at zeroth-order (i.e.,
treated in the classical Bayesian manner). In the numerical examples
of this paper, linear expansions are used for the embedded $\lambda$
components. A demonstration of choosing the embedding components will
be shown in the numerical results.  The full set of parameters to be
calibrated is grouped together via the notation $\talpha \equiv
(\lambda, \alpha)$.  The model calibration problem thus involves
finding the posterior distribution on $\talpha$ via Bayes' theorem
\begin{align}
  p(\talpha | D) = \frac{p(D|\talpha) p(\talpha)}{p(D)},
          \label{e:posterior} 
\end{align}
where $D = \left\{g_k \right\}_{k=1}^K$ is the set of $K$ calibration
data points (which are the high-fidelity model evaluations in this
case), $p(\talpha)$ is the prior distribution, $p(D|\talpha)$ is the
likelihood function, and $p(D)$ is the evidence. The prior and
posterior represent our states of knowledge about the uncertainty in
the parameters $\talpha$ before and after the data set $D$ is
assimilated.  To facilitate Bayesian inference and obtain the
posterior in a practical manner, we will further develop the
likelihood model below.  Once the posterior is characterized, it can
be subsequently propagated through the low-fidelity model to obtain
posterior predictive distributions on the desired QOIs---that is,
predictions that account for both model error and parameter
uncertainty.

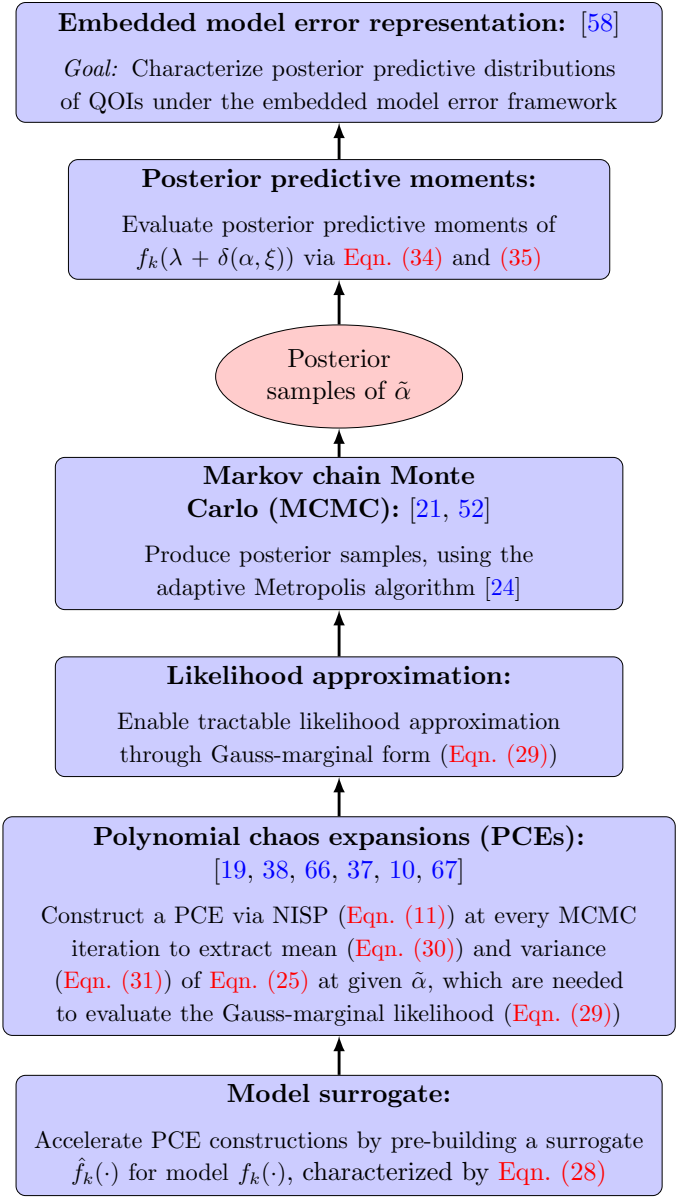
\begin{figure}
  \centering
  \resizebox{0.56\textwidth}{!}{

        \tikzset{      dimen/.style={<->,>=latex,very thick,every rectangle node/.style={midway,font=\footnotesize}},
    }
    
    \tikzstyle{block} = [rectangle, draw, fill=blue!20, 
      text width=20em, text centered, rounded corners, minimum height=4em]
    \tikzstyle{cloud} = [draw, ellipse, text centered, fill=red!20, node distance=3cm,
      minimum height=2em]

    \begin{tikzpicture}[node distance = 7.em, auto]

            \node [block, text width=24em] (merr)
      {\textbf{Embedded model error
          representation:} \cite{Sargsyan2015}\\\vspace{0.5em}{\small
          \textit{Goal:} Characterize posterior predictive
          distributions of QOIs under the embedded model error
          framework}};
                                                            \node [block, below of=merr, text width=20em, node distance=6em]
      (predictive_moments) {\textbf{Posterior predictive
          moments:}\\\vspace{0.5em}{\small 
        Evaluate posterior predictive moments of
        $f_k(\lambda + \delta(\alpha,\xi))$ via
        \cref{e:predictive_mean,e:predictive_var}}};
      \node [cloud, below of=predictive_moments, text width=6em, node distance=6em]
      (posterior) {Posterior samples of $\talpha$};
                  \node [block, below of=posterior, text width=21em, node
        distance=6em] (MCMC) {\textbf{Markov 
        chain Monte Carlo (MCMC):} \cite{Gilks1996, Roberts2004}\\\vspace{0.5em}{\small
          Produce posterior samples, using the adaptive Metropolis
          algorithm~\cite{Haario2001}}};
      \node [block, below of=MCMC, text width=21em, node distance=7em] (likelihood)
            {\textbf{Likelihood approximation:}\\\vspace{0.5em}{\small
          Enable tractable likelihood approximation through
          Gauss-marginal form (\cref{e:gausmarg})}};
      \node [block, below of=likelihood, text width=25em, node distance=8em] (PCE)
            {\textbf{Polynomial chaos expansions
                (PCEs):}\\\cite{Ghanem1991, Najm2009, Xiu2009, 
            LeMaitre2010, Ernst2012, Xiu2002}\\\vspace{0.5em}{\small
          Construct a PCE via NISP (\cref{e:NISP}) at every MCMC
          iteration to extract mean 
          (\cref{e:gausmarg_mean}) 
          and variance (\cref{e:gausmarg_var}) of
            \cref{e:embed_discretize} at given $\talpha$,
            which are needed to evaluate the 
            Gauss-marginal likelihood (\cref{e:gausmarg})}};
      \node [block, below of=PCE, text width=24em, node distance=8em] (surrogate)
            {\textbf{Model surrogate:}\\\vspace{0.5em}{\small
          Accelerate PCE constructions by pre-building a surrogate
          $\hf_k(\cdot)$ for 
          model $f_k(\cdot)$}, characterized by
              \cref{e:embed_with_surrogate}};

            \draw [dimen,<-] (merr) -- (predictive_moments);
      \draw [dimen,<-] (predictive_moments) -- (posterior);
      \draw [dimen,<-] (posterior) -- (MCMC);
      \draw [dimen,<-] (MCMC) -- (likelihood);
      \draw [dimen,<-] (likelihood) -- (PCE);
      \draw [dimen,<-] (PCE) -- (surrogate);

    \end{tikzpicture}
  }
  \caption{Summary of the computational concepts and tools used for
    the embedded model error representation.}
  \label{f:merr_flow}
\end{figure}

\Cref{f:merr_flow} shows a summary of the computational concepts and
tools used for the embedded model error representation, and may serve
as a useful reference for readers as they are introduced below.

\subsection{Surrogate for low-fidelity model}

Under this framework, the high-fidelity model only needs to be
evaluated to generate data for Bayesian inference; this typically
involves a small number of evaluations.
The low-fidelity model, however, needs to be run as many times as is
required to perform Bayesian inference in characterizing the posterior
$p(\talpha|D)$; this entails a much larger number of evaluations in
comparison. When the low-fidelity model evaluations are still
expensive, a surrogate model is needed. We proceed to build a
surrogate using Legendre polynomials for each QOI involved in either
calibration or prediction, $f_k(\cdot)\approx \hf_k(\cdot)$, as
functions of the overall input argument
$(\lambda+\delta(\alpha,\xi))$, to replace the low-fidelity model.
The approximation error is represented with an additive Gaussian form,
and so \cref{e:embed_discretize} becomes
\begin{align}
  q_k = f_k(\lambda + \delta(\alpha,\xi)) = \hf_k(\lambda +
  \delta(\alpha, \xi)) + \epsilon_k,\label{e:embed_with_surrogate}
\end{align}
where $\hf$ is the surrogate to the low-fidelity model, and $\epsilon$
encapsulates the error of the surrogate with respect to the
low-fidelity model. In this study, it is assumed $\epsilon_k\iid
\CN\(0,\sigma^2_{k,\mathrm{LOO}}\)$ and independent of the surrogate
model input (but depends on $k$). The variance terms
$\sigma^2_{k,\mathrm{LOO}}$ are the leave-one-out cross-validation
errors from the linear regression systems used for constructing the
surrogates, and can be computed analytically and quickly
(e.g.,~\cite{Christensen2011}).  We emphasize that
$\delta(\alpha,\xi)$ is still representing the model discrepancy
between the high-fidelity and low-fidelity models, \textit{not}
between the high-fidelity and the surrogate models.

\subsection{Likelihood approximation}

We characterize the posterior $p(\talpha|D)$ via Markov chain Monte
Carlo sampling~\cite{Gilks1996, Roberts2004}, specifically using the
adaptive Metropolis algorithm~\cite{Haario2001}.  MCMC requires the
evaluation of the prior $p(\talpha)$ and likelihood $p(D|\talpha)$ at
every iteration. Uniform prior distributions are adopted to allow,
together with Legendre-uniform PCEs, better control on the range of
the physical parameters.  Direct evaluation of the likelihood is
intractable, since $p(D|\talpha)$ does not have a closed form and
would require either kernel density estimation or numerical
integration, both of which are very expensive. Furthermore, the
likelihood often involves highly nonlinear and near-degenerate
features (in fact, it is fully degenerate when data noise is
absent~\cite{Sargsyan2015}).  These challenges motivate us to seek
alternative forms that approximate the likelihood in a computationally
feasible manner.

Sargsyan~\textit{et al.}~\cite{Sargsyan2015} suggested several options
based on the assumption of conditional independence between data
points. In this study, we adopt the Gaussian approximation to a
marginalized likelihood:
\begin{align}
  p(D|\talpha) \approx L_{G}(\talpha) = \frac{1}{(2\pi)^{\frac{N}{2}}}
  \prod_{k=1}^N \frac{1}{\sigma_k(\talpha)}
  \exp\[-\frac{(\mu_k(\talpha)-g_k)^2}{2\sigma_k^2(\talpha)}\],
  \label{e:gausmarg}
\end{align}
where
\begin{align}
  \mu_k(\talpha) = \EE\[\hf_k(\lambda+\delta(\alpha,\xi)) +
  \epsilon_k\] = \EE_{\xi}\[\hf_k(\lambda+\delta(\alpha,\xi))\]
  \label{e:gausmarg_mean}
\end{align}
and
\begin{align}
  \sigma_k^2(\talpha) =
  \Var\[\hf_k(\lambda+\delta(\alpha,\xi))+\epsilon_k\] =
  \Var_{\xi}\[\hf_k(\lambda+\delta(\alpha,\xi))\] + \sigma^2_{k,\mathrm{LOO}}
  \label{e:gausmarg_var}
\end{align}
are the mean and variance of the low-fidelity model at fixed
$\talpha=(\lambda,\alpha)$.  We estimate these moments by constructing
a PCE for the outputs from propagating the PCE of the input argument
in \cref{e:lambda_PCE}:
\begin{align}
  \hf_k(\lambda+\delta(\alpha,\xi)) = \hf_k\(\lambda + \sum_{\beta\neq
  0}
  \alpha_{\beta} \Psi_{\beta}(\xi)\) \approx \sum_{\beta}
  \hf_{k,\beta}(\talpha)\Psi_{\beta}(\xi).\label{e:surrogate_PCE}
\end{align}
This can be done using the NISP method described in \cref{e:NISP}
together with quadrature integration. The moments can then be
extracted from the expansion coefficients as \begin{align}
 \mu_k(\talpha) \approx  \hf_{k,0}(\talpha) \qquad\qquad \textrm{ and } \qquad\qquad \sigma_k^2(\talpha) \approx \sum_{\beta\neq 0}
  \hf_{k,\beta}^2(\talpha).
 \label{e:moments_PCE}
\end{align}
When using a polynomial surrogate, and if a linear input PCE is
employed, NISP provides exact equality in \cref{e:surrogate_PCE}.

\subsection{Posterior predictive}
\label{ss:prediction}

Once posterior samples are generated from MCMC, prediction samples for
desired QOIs can be produced by evaluating
\cref{e:embed_with_surrogate} at each of these $\talpha$ samples
together with randomly generated values of $\xi$. Instead of
characterizing the posterior predictive distribution using samples,
however, we opt to evaluate its first two moments (mean and variance)
since they can be estimated easily using existing analytical moment
information with respect to $\xi$ (i.e., $\mu_k$ and $\sigma_k$). The
posterior predictive mean is
\begin{align} 
  \EE\[q_k\]=\EE\[\hf_k(\lambda + \delta(\alpha, \xi))+ \epsilon_k\]
  = \EE_{\talpha}\[\hf_k(\lambda + \delta(\alpha, \xi))\] =
  \EE_{\talpha}\[\mu_k(\talpha)\]
  \label{e:predictive_mean}
\end{align}
and variance
\begin{align}
  \Var\[q_k\]&=\EE_{\talpha}\left[\Var_{\xi}\[\hf_k(\lambda
      + \delta(\alpha, \xi))\]\right]
  +\Var_{\talpha}\left[\EE_{\xi}\[\hf_k(\lambda
      + \delta(\alpha, \xi))
      \]\right]
  +\Var\[\epsilon_k\] \nonumber\\[0.7em] &=
  \underbrace{\EE_{\talpha}\left[\sigma_k^2(\lambda,\alpha)\right]}_{\textrm{model
      error}} +\underbrace{\Var_{\talpha}\[\mu_k(\lambda,
    \alpha)\]}_{\textrm{posterior uncertainty}}
  +\underbrace{\sigma^2_{k,\mathrm{LOO}}}_{\textrm{surrogate
      error}},\label{e:predictive_var}
\end{align}
where all ${\EE}_{\talpha}$ and $\Var_{\talpha}$ are with respect to
the posterior, and are computed by standard estimators using posterior
samples from MCMC. The decomposition of the variance invokes the law
of total variance, and allows us to attribute the overall predictive
variance to components due to model error, parameter posterior, and
surrogate error. These quantities are estimated by applying the
surrogate in \cref{e:surrogate_PCE} to the MCMC samples.

\section{Numerical Results}
\label{s:results}

\subsection{Global sensitivity analysis}

We demonstrate our GSA machinery through a study with 24 input
parameters shown in \cref{t:GSA_params}. Within this set, all scalar
parameters (other than the wall temperature $T_{w}$ boundary
condition) are endowed with independent uniform distributions across
ranges suggested from experts in the field.\footnote{We acknowledge
  that correlation between these parameters exists due to the
  underlying physics, but attempting to discover the correlation
  information (e.g., through a Bayesian inference problem) is beyond
  the scope of this paper. The independent uniform distributions can
  be viewed as maximum entropy densities based on specified parameter
  ranges.  The subsequent GSA results therefore correspond to these
  distributions, and would be different if correlation information
  were injected.}  $T_{w}$ is a function of the continuous streamwise
coordinate $x/d$, and hence is a random field (RF). In this study, we
use the normalized spatial coordinates $x/d$, $y/d$, and $z/d$ for
convenience, where $d=3.175$ mm is the diameter of the
injector. $T_{w}$ is represented using the Karhunen-Lo\`{e}ve
expansion (KLE) (see e.g., Ghanem and Spanos~\cite{Ghanem1991}), which
is built employing the eigenstructure of the covariance function of
the RF to achieve an optimal representation. We employ a Gaussian RF
with a square exponential covariance structure along with a
correlation length that is similar to the largest turbulent eddies
(i.e., the size of the oxidizer inlet). The mean temperature profile
is constructed by averaging temperature profile results from a small
set of separate adiabatic simulations. The correlation length employed
leads to a rapid decay in characteristic-mode amplitudes, allowing us
to capture about 90\% of the total variance of this RF with only a 10
dimensional KLE. The wall temperature is further assumed to be
constant in time.

\begin{table}[htb]
\begin{center}
\begin{tabular}{cll}
\hline
Parameter &  Range  & Description \\
\hline 
\multicolumn{3}{c}{Inlet boundary conditions}\\
\hline
$p_{0}$ & [1.406, 1.554] MPa & Stagnation pressure   \\
$T_{0}$ & [1472.5, 1627.5] K & Stagnation temperature  \\
$M_0$ & [2.259, 2.761] & Mach number  \\
$\delta_{a}$ & [2, 6] mm & Boundary layer thickness \\
$I_{i}$ & [0, 0.05] & Turbulence intensity magnitude  \\
$L_{i}$ & [0, 8] mm & Turbulence length scale  \\
\\
\hline
\multicolumn{3}{c}{Fuel inflow boundary conditions}\\
\hline
$\dot{m}_f$ & [6.633, 8.107] $\times 10^{-3}$ kg/s & Mass flux   \\
$T_{f}$ & [285, 315] K & Static temperature \\
$M_f$ & [0.95, 1.05] & Mach number \\
$I_{f}$ & [0, 0.05] & Turbulence intensity magnitude \\
$L_{f}$ & [0, 1] mm & Turbulence length scale  \\
\\
\hline 
\multicolumn{3}{c}{Turbulence model parameters} \\ \hline    
$C_R$ & [0.01, 0.06]  & Modified Smagorinsky constant  \\
$Pr_t$ & [0.5, 1.7] & Turbulent Prandtl number \\
$Sc_t$ & [0.5, 1.7] & Turbulent Schmidt number \\
\\
\hline 
\multicolumn{3}{c}{ Wall boundary conditions}    \\  \hline
$T_{w}$  & Expansion in 10 params & Wall temperature represented via \\
& of $\CN(0,1)$ & Karhunen-Lo\`{e}ve expansion \\ 
\hline
\end{tabular}
\end{center}
\caption{Uncertain parameters for the GSA example.}
\label{t:GSA_params}
\end{table}

For the ML and MF representations, we consider four combinations of
grid resolutions and model fidelity: grid resolutions $d/8$ and
$d/16$, for 2D and 3D simulations. A grid resolution of $d/8$ means
the injector diameter $d=3.175$ mm is discretized by 8 grid
cells. Hence, $d/8$ depicts a ``coarse'' grid, and $d/16$ a ``fine''
grid. In this study, we aim to construct the following two telescopic
PCEs:
\begin{align}
    \hat{f}_{2D,d/16}(\lambda) &= \hat{f}_{2D,d/8}(\lambda) +
    \hat{f}_{\Delta_{2D,d/16-2D,d/8}}(\lambda)\label{e:ml_scramjet}
    \\ \hat{f}_{3D,d/8}(\lambda) &= \hat{f}_{2D,d/8}(\lambda) +
    \hat{f}_{\Delta_{3D,d/8-2D,d/8}}(\lambda)\label{e:mf_scramjet}.
\end{align}
The first is an exercise of ML, while the second is MF.  More
sophisticated PCEs representing the 3D $d/16$ model are also possible,
but we do not attempt them here due to the limited number of 3D $d/16$
runs.

\begin{figure}[htb]
  \centering
  \includegraphics[width=0.32\textwidth]{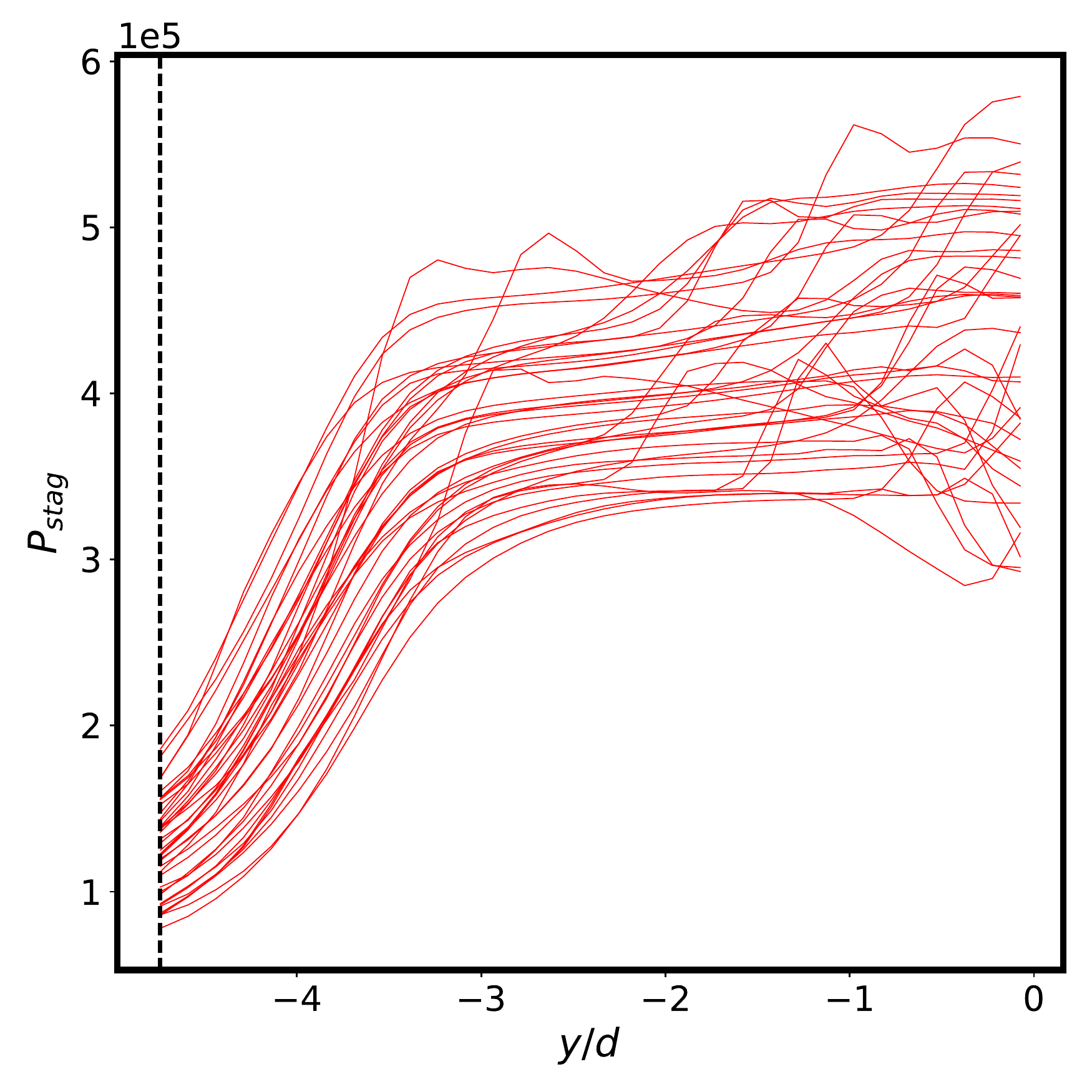}
  \includegraphics[width=0.32\textwidth]{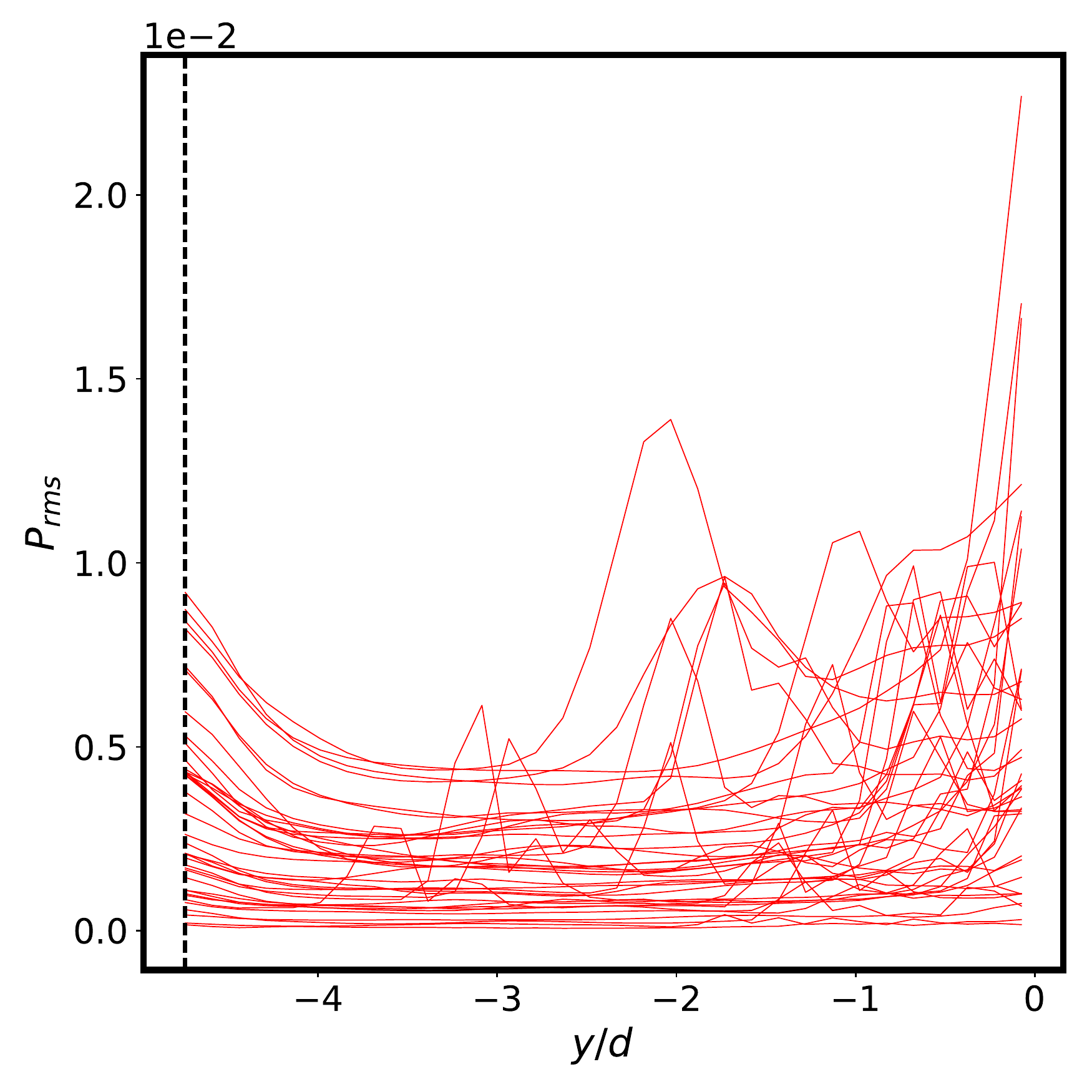}
  \includegraphics[width=0.32\textwidth]{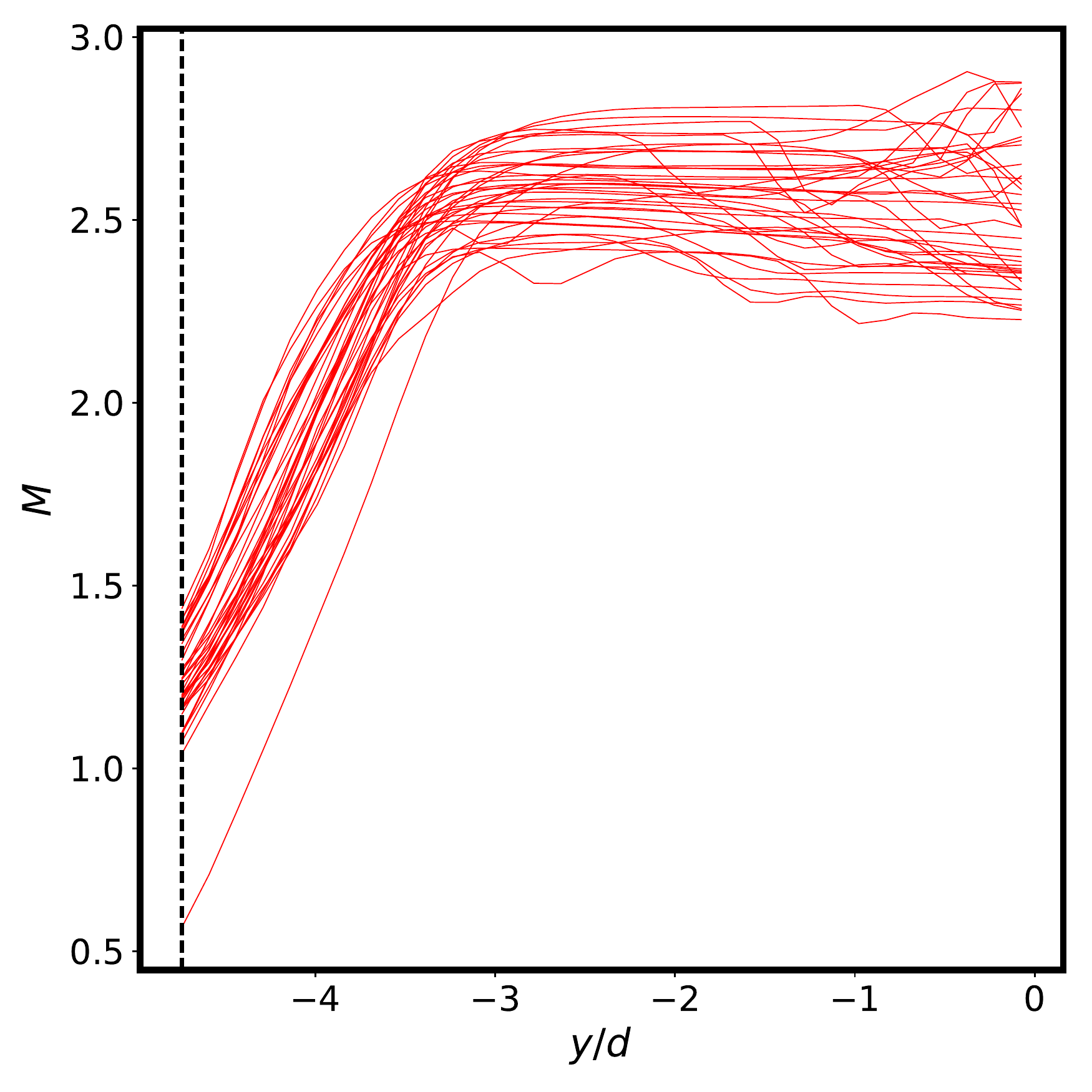}
  \includegraphics[width=0.32\textwidth]{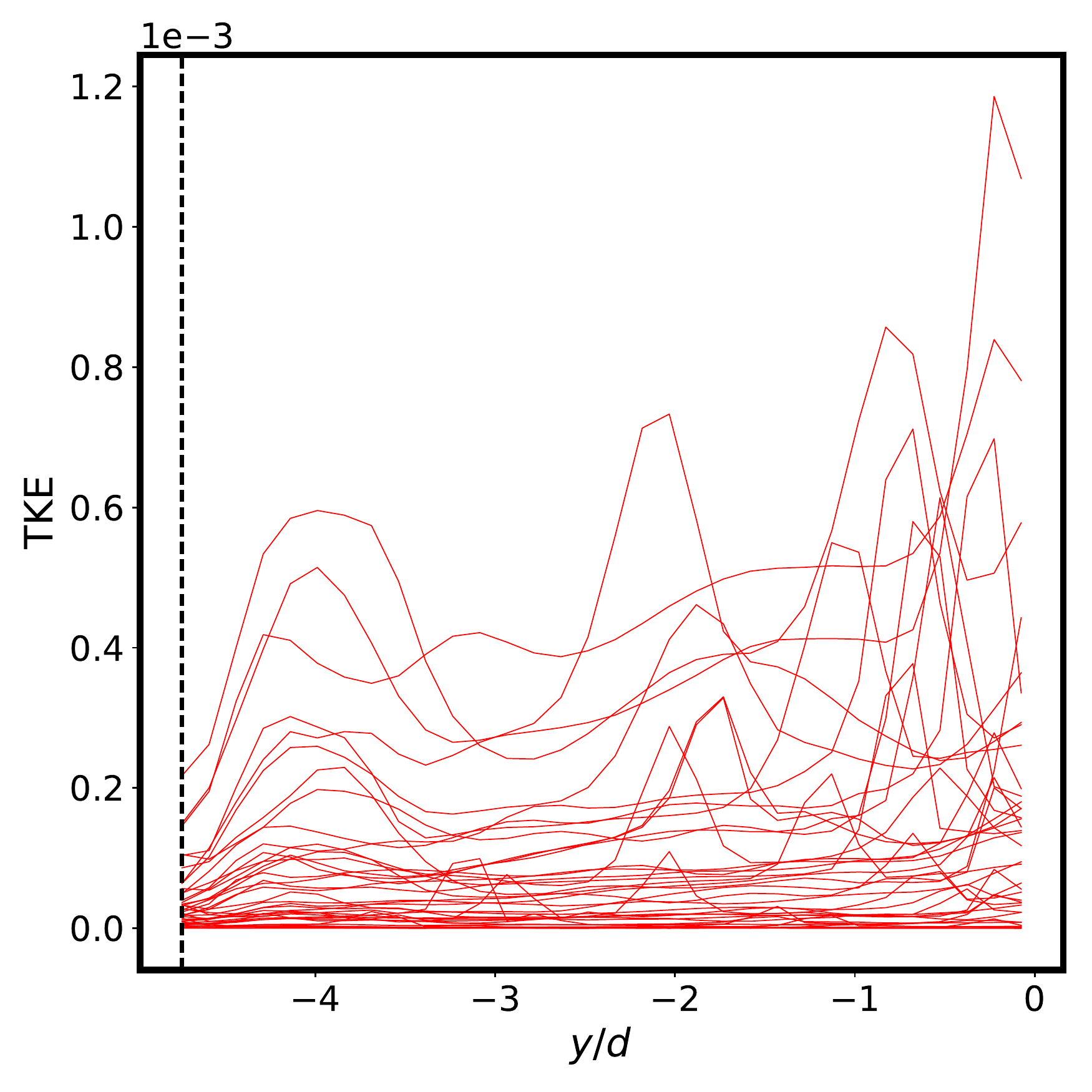}
  \includegraphics[width=0.32\textwidth]{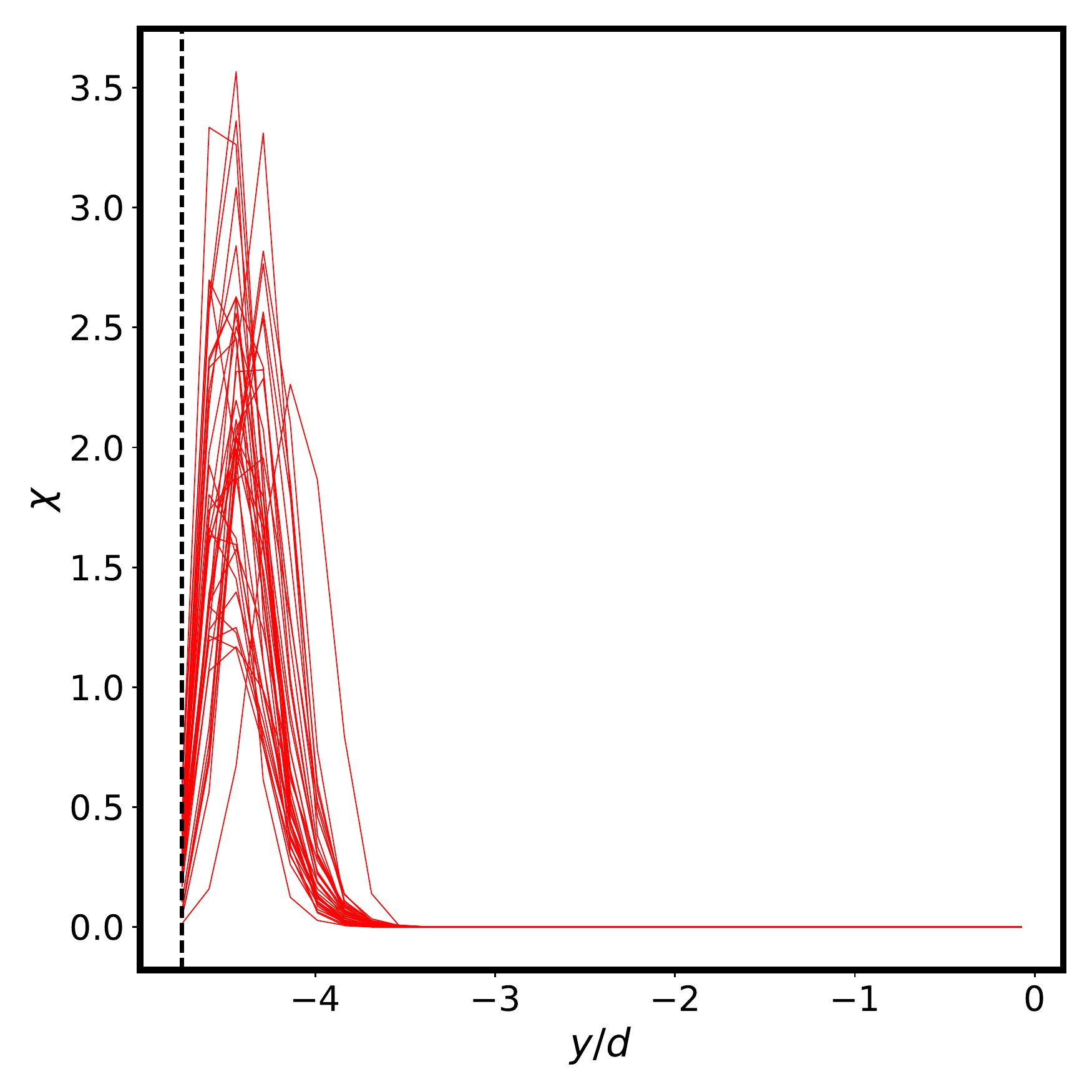}
  \caption{Profiles of five targeted observables in $y/d$, at fixed
    $x/d=100$, and averaged over time for selected 2D $d/8$ runs of
    different parameter settings in the GSA study. }
  \label{f:mlmf_profile_examples}
\end{figure}

Five output observables are studied: stagnation pressure $P_{stag}$,
root-mean-square (rms) static pressure $P_{rms}$, Mach number $M$,
turbulent kinetic energy TKE, and scalar dissipation rate
$\chi$. Examples of profiles for these observables across the
wall-normal direction $y/d$, at fixed streamwise direction $x/d=100$,
and averaged over time are shown in \cref{f:mlmf_profile_examples} for
2D $d/8$ under various parameter settings. In these plots, each red
line is the profile of an independent simulation. The left side
represents the lower wall of the chamber, with the dotted vertical
line depicting the location of the wall; the right side represents the
symmetry line of the chamber center.  Effects of the boundary layer
can be clearly seen through $P_{stag}$, $M$, and $\chi$.

Our QOIs are the \textit{mean} statistics of these five variables
across $y/d$ and at fixed $x/d=100$; all QOIs are also
time-averaged.\footnote{\label{f:time_avg_GSA}The variation due to
  time averaging depends on the location of averaging window (i.e.,
  whether steady-state is reached), the inherent variation of the QOI,
  and the number of samples used for averaging. We perform the
  averaging over a window after transient behavior has sufficiently
  subsided, and detailed investigations on select runs indicate the
  overall time averaging variations is negligible compared to the
  variation of QOIs over the parameter space. We thus do not include
  them in the numerical results presented.}  Results for 3D
simulations are taken at the centerline of the spanwise
direction. Sample allocation across different models is calculated
using MLMF MC, which minimizes the aggregate variance of MC estimators
that correspond to each of the five QOIs. The number of runs for each
model can be found in \cref{t:MLMFMC}, where the desired general trend
of fewer expensive simulations and more inexpensive simulations is
evident (note that while 3D $d/16$ runs are not used in the PCE
constructions above, they are still part of the MLMF MC allocation
algorithm).

\begin{table}[htb]
\begin{center}
\begin{tabular}{l|cc}
\hline
&  2D  & 3D \\\hline
$d/8$ (coarse grid) & 1939 & 46 \\
$d/16$ (fine grid) & 79 & 11 \\
\hline 
\end{tabular}
\end{center}
\caption{Number of samples from the MLMF MC allocation algorithm
  (converged runs only). }
\label{t:MLMFMC}
\end{table}

The PCEs in \cref{e:ml_scramjet,e:mf_scramjet} are built for each QOI
separately, using all available samples. GPSR is used to find a sparse
PCE for each term, which are then combined together before a final
relative thresholding (i.e., with respect to the coefficient of
largest magnitude in that expansion) of $10^{-3}$ is applied.
For PCEs with total-order of degree 3, GPSR is able to downselect from
a full set of 2925 basis terms to 187, 1336, 1676, 663, 2302 for the
five QOIs in ML, and 200, 96, 308, 51, 352 in MF. The larger number of
terms in ML is due to fewer available samples and a low
signal-to-noise ratio for the difference between fine and coarse grid
results.

\begin{figure}[htb]
  \centering
  \includegraphics[width=0.49\textwidth]{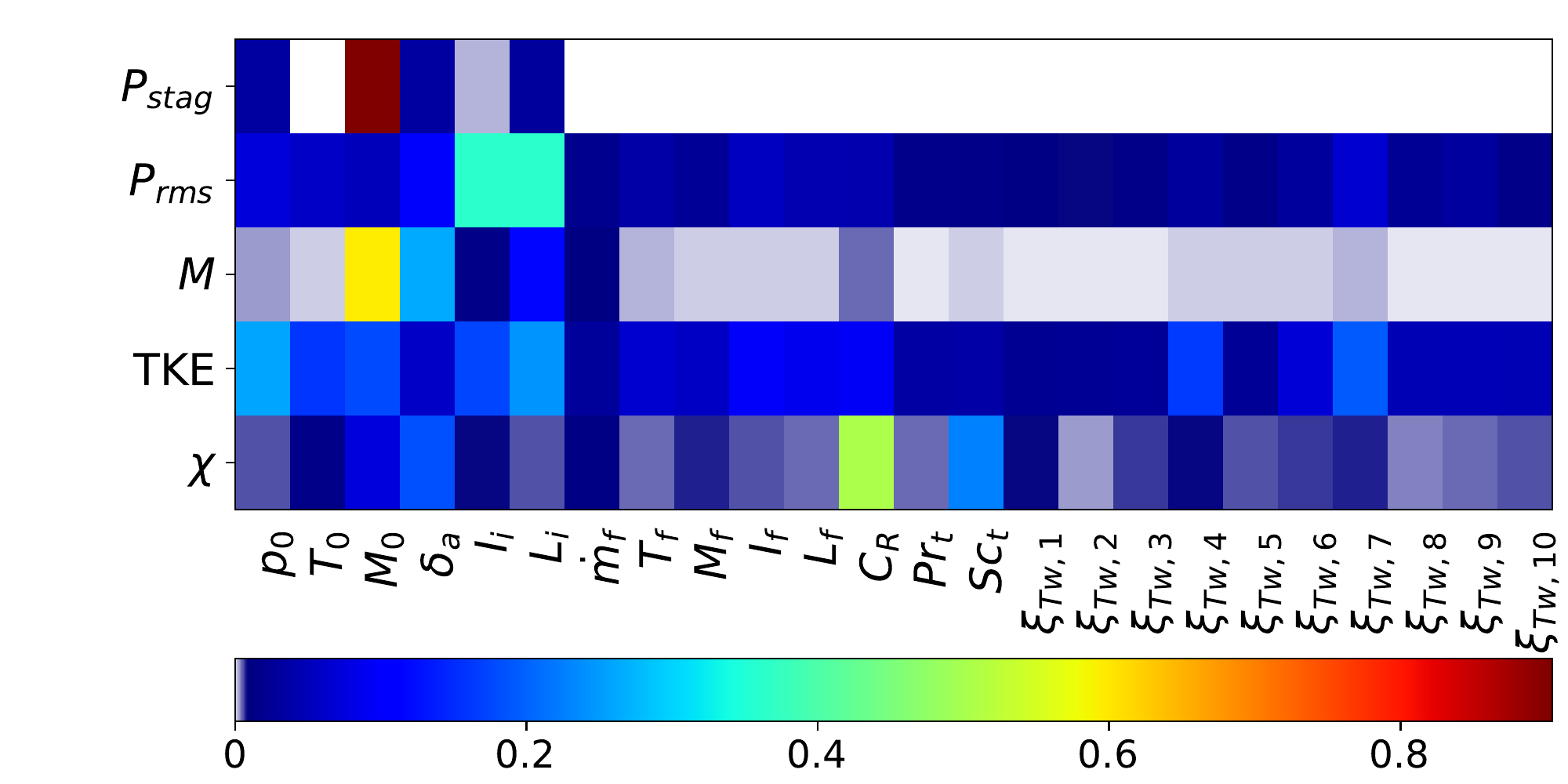}
  \includegraphics[width=0.49\textwidth]{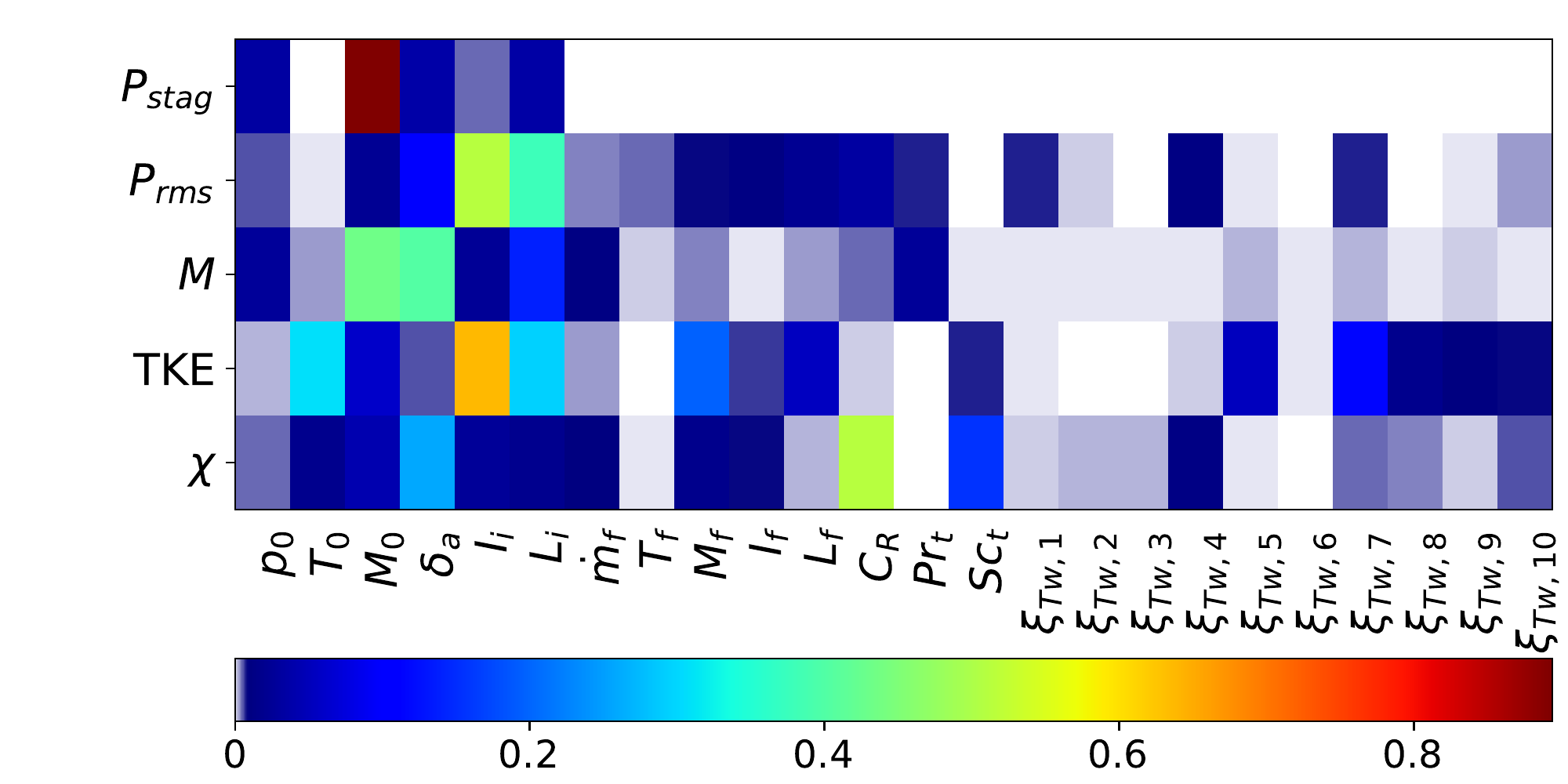}
  \caption{Total effect sensitivity indices for the ML (left) and MF
    (right) expansions corresponding to
    \cref{e:ml_scramjet,e:mf_scramjet}, respectively.  }
  \label{f:mlmf_gsa}
\end{figure}

The total effect sensitivity indices are plotted in \cref{f:mlmf_gsa}
for the ML and MF expansions.  The rows depict the five QOIs (only the
variable names are displayed in the figures, with the QOIs being these
variables averaged over time and $y/d$, and at $x/d=100$) and the
columns correspond to the 24 input parameters. $\xi_{Tw,i}$ denotes
the $i$th KLE term for the wall temperature random field. Red spectrum
indicates high sensitivity, and blue indicates low sensitivity (with
grey-white being near-zero sensitivity). Sensitivity is relative
within each QOI, thus one should not compare their values across
different rows.  Since the overall QOIs represented by the two
expansions correspond to different fidelity and grid resolutions, one
should not expect identical results (even under infinite samples),
although it is reasonable to observe similar qualitative behavior. For
both expansions, the most sensitive inputs tend to be those related to
inflow conditions---$M_0$, $\delta_a$, $L_i$, and $I_i$ have high
sensitivity for one or more of the five QOIs. $T_0$ and $p_0$ also
appear to be influential for the mean TKE, while $C_R$ is quite
important for the mean scalar dissipation rate $\chi$. Parameters
corresponding to the wall temperature KLE have small contributions for
most QOIs with all ten modes contributing under 5\%
variance. Exceptions are observed for mean $P_{rms}$ receiving 25\% in
the ML expansion, and mean TKE receiving 67\% and 21\% in the ML and
MF expansions, respectively.  Overall, these observations are
consistent with our physical intuition: since the current
jet-in-crossflow test problem does not involve combustion, one would
expect the bulk inflow conditions to dominate the impact on the QOI
behavior.  In future work involving the full HDCR geometry with
combustion enabled, we expect to reveal different sensitivity trends
under new interactions that would be otherwise nonintuitive and
non-obvious.

\subsection{Model error}

We demonstrate the embedded representation of model error via two
examples.  The first involves 3D LES on a grid with resolution
$d/8$. The high-fidelity model uses a dynamic treatment of Smagorinsky
turbulent characterization, where the Smagorinsky constant and
turbulent Prandtl and Schmidt numbers are calculated locally at every
grid point. The low-fidelity model employs a static treatment, where
those parameters are fixed globally across the entire grid by the
user. The static version provides about 30\% saving of computational
time.
The second example calibrates a low-fidelity 2D model from evaluations
of high-fidelity 3D simulations, and both models are simulated on the
$d/8$ grid with the static Smagorinsky treatment. Even on this coarse
grid, a 2D run requires two orders of magnitude less computational
time compared to its 3D counterpart, and finer meshes would provide
even greater cost savings. This is thus a realistic situation with
strong tradeoff between accuracy and cost. In both cases, we would
like to quantify the uncertainty due to model error when the less
expensive low-fidelity model is used.

\subsubsection{Static versus dynamic Smagorinsky}

For the static model, we augment the modified Smagorinsky constant
$\lambda = C_R$ with an additive model error representation. It is
endowed with a uniform prior $C_R\sim \CU(0.005, 0.08)$, which is
selected based on existing literature and preliminary simulation
tests. The turbulent Prandtl and Schmidt numbers ($Pr_t$ and $Sc_t$)
are both fixed at a nominal value of $0.7$. The calibration data are
chosen to be the discretized profile of TKE along $y/d$, and at fixed
$x/d=100$ and spanwise centerline, for a total of 31 nodes. All QOIs,
for both calibration and prediction, are
time-averaged.\footnote{\label{f:time_avg_merr}Similar to
  \cref{f:time_avg_GSA} in the GSA part, investigations on select
  cases indicate that variation due to time averaging is negligible
  compared to uncertainty contributions from model error and parameter
  posterior. We thus do not include them in the numerical results
  presented.} Other choices of calibration QOIs are certainly
possible, and ideally we would calibrate using the same QOIs for which
we are interested in making predictions.  We choose TKE here for
demonstration.  An interesting topic of future research is when
calibration data are not available for predictive QOIs. Optimal
experimental design methods can then be used to help choose from the
available calibration QOIs those that are most useful for the
predictive QOIs.

We set the additive term $\delta(\alpha,\xi)$ (from
\cref{e:embed_with_surrogate}) to be first-order Legendre-Uniform PCE
in $C_R$, i.e., model error representation is embedded in $C_R$. The
low-fidelity model surrogate, represented as a third-order Legendre
polynomial, is built via regression using $9$ training samples, i.e.,
$9$ evaluations of the low-fidelity model at different $C_R$ values.
The PCEs used for the likelihood and posterior predictive are
third-order Legendre-Uniform, and integration over $\xi$ is performed
using the 4-point Gaussian quadrature rule in each dimension. MCMC is
run for $10^5$ samples, with 50\% burn-in and thinning of every 100
samples to improve mixing.

\begin{figure}[htb]
  \centering
  \includegraphics[width=0.45\textwidth]{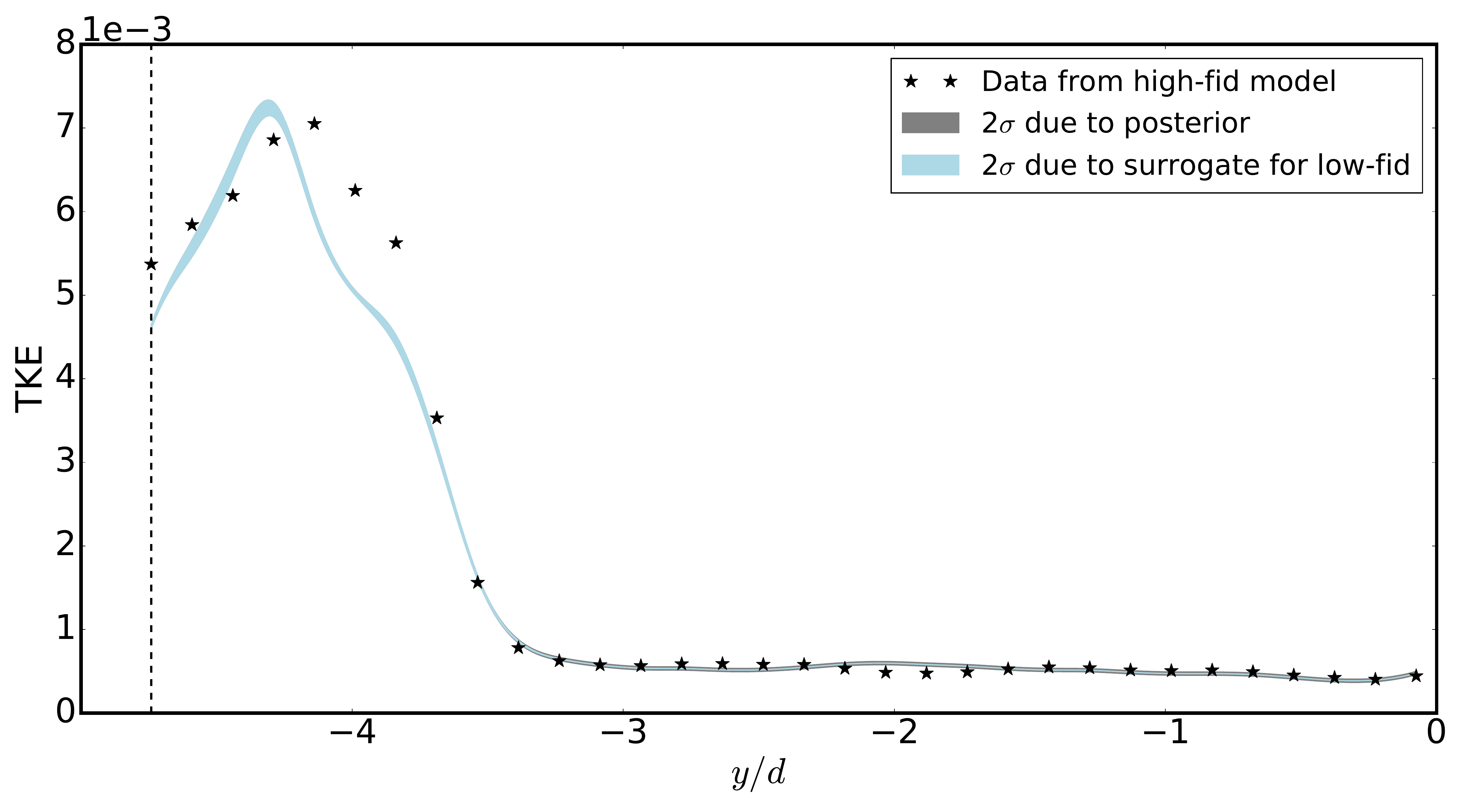}
  \includegraphics[width=0.45\textwidth]{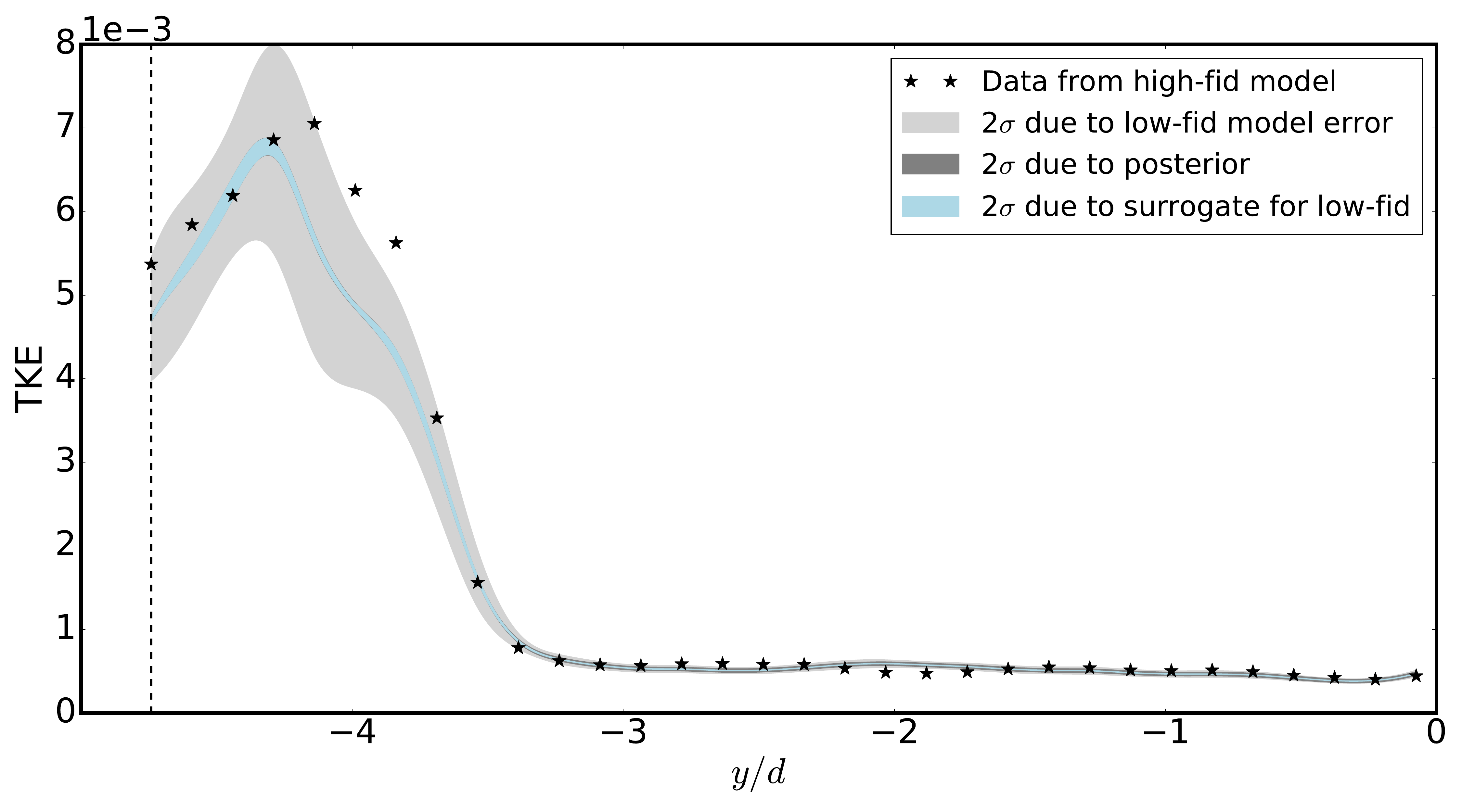}
  \includegraphics[width=0.45\textwidth]{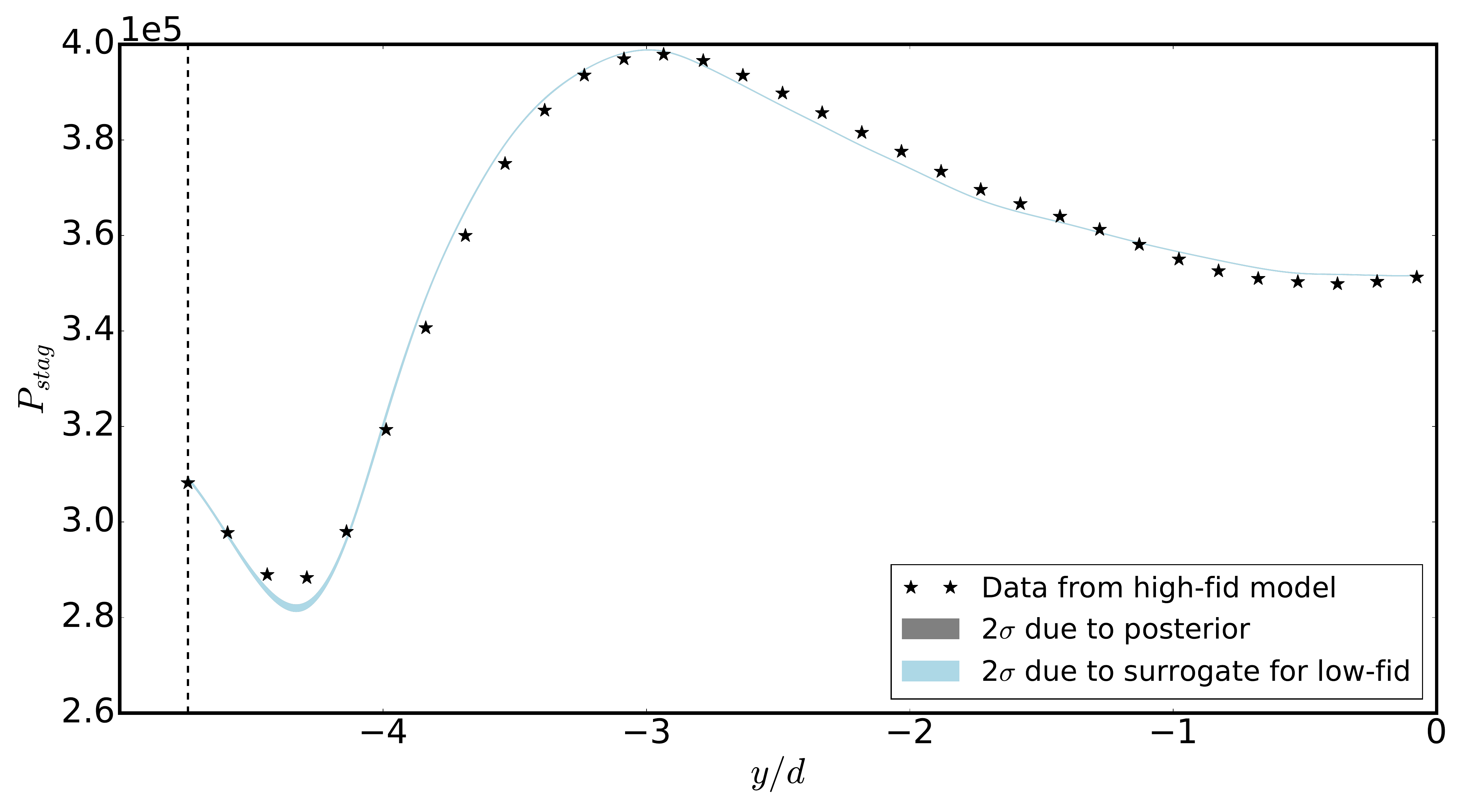}
  \includegraphics[width=0.45\textwidth]{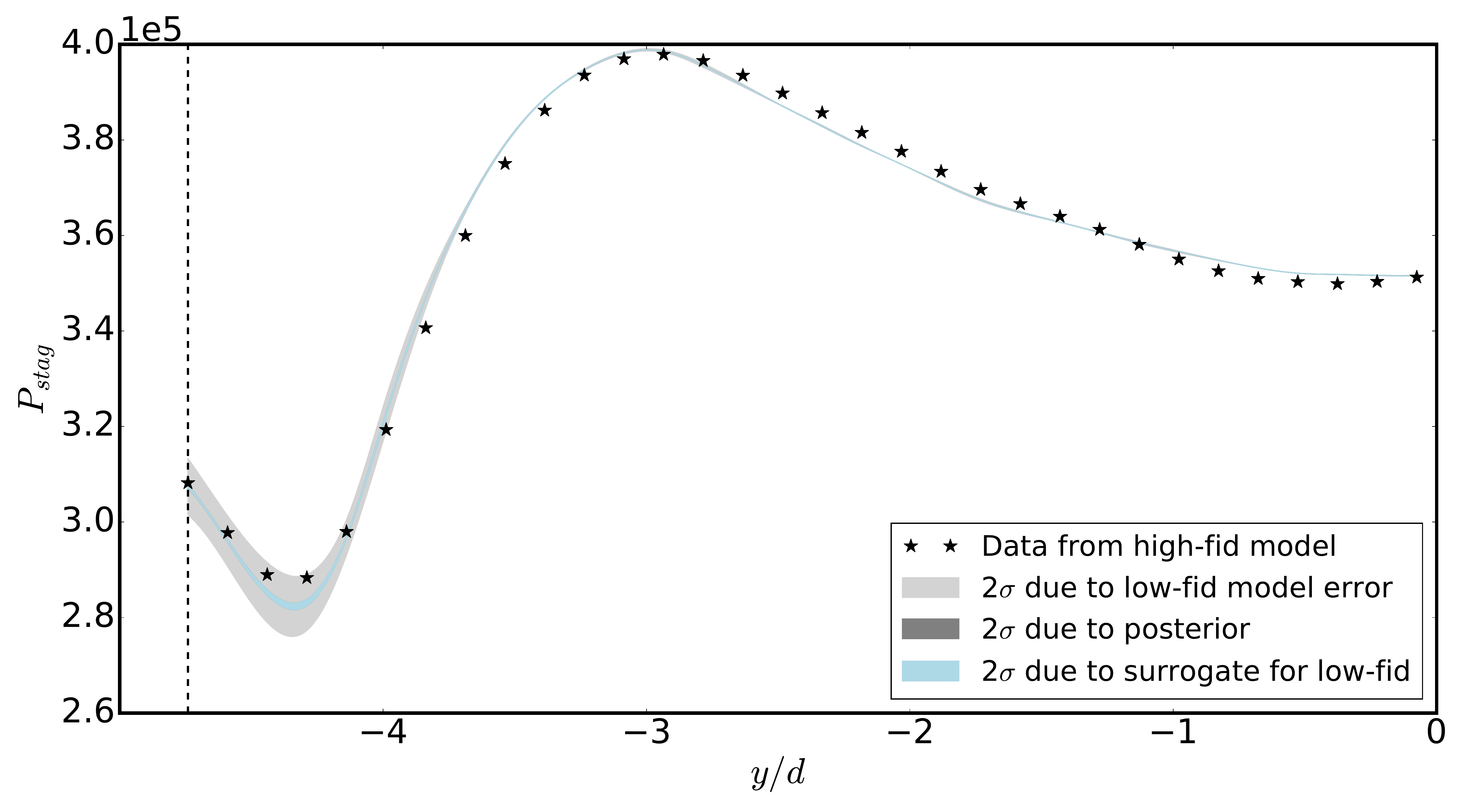}
  \caption{Posterior predictive distributions for TKE and $P_{stag}$
    profiles from the static Smagorinsky model without model error
    treatment (left column) and with embedded model error
    representation (right column). }
  \label{f:static_vs_dynamic_fullprofiles}
\end{figure}

\Cref{f:static_vs_dynamic_fullprofiles} depicts the static Smagorinsky
model posterior predictive distributions for the profiles of TKE and
$P_{stag}$. The TKE profile constitutes the data set used for
calibration, while the $P_{stag}$ profile is extrapolatory.  The left
column displays classical Bayesian inference results, while the right
column contains results with model error representation.  The black
dots are the true high-fidelity evaluations, and the light grey, dark
grey, and blue bands represent $\pm 2$ standard deviations due to
model error, posterior, and low-fidelity surrogate uncertainty,
respectively, as broken down in \cref{e:predictive_var}. In this case,
the data set is overall quite informative, leading to very narrow
posterior distributions (dark grey band) in all figures.  The
classical inference results in the left column lead us to be
overconfident in predictive results that do not match well with the
high-fidelity model in many regions.  The strength of the model error
representation is evident in the right column, as the light grey bands
allow much better capturing of the model-to-model discrepancy, and
present a better indication of our loss in model accuracy. In this
example, the model error is characterized well for the extrapolatory
QOIs ($P_{stag}$ profile, bottom-right figure). This may not always be
the case, since the extrapolation of $\delta(\cdot)$ to QOIs outside
those used for calibration may be inadequate, and there may be
differences between high- and low-fidelity models that cannot be
captured solely from parameter embedding. We will illustrate these
challenges and limitations in the next example. Finally, we emphasize
that all realizations generated from the posterior predictive
distributions under this framework automatically satisfy the governing
equations of the low-fidelity model.

\subsubsection{2D versus 3D}

In the second example, we calibrate a 2D model using data from 3D
simulations.  The parameters for the 2D model are $\lambda = (C_R,
Pr_t^{-1}, Sc_t^{-1}, I_i, I_r, L_i)$ endowed with uniform priors (see
\cref{t:twoD_vs_threeD_params} for their definitions and prior
ranges). Note that, for this example, we target the \textit{inverse}
Prandtl and Schmidt numbers instead of the non-inverted version used
in the GSA cases.
The 3D calibration data are generated at a fixed condition of
$\lambda_{3D}^*=(0.0297,1/0.703, 1/0.703,0.05,1.0, 0.00423)$.  The
calibration data are chosen to be the discretized profile of scalar
dissipation rate $\chi$ along $y/d$, at fixed $x/d=88$ and spanwise
centerline, for a total of 31 nodes.

\begin{figure}[htb]
  \centering
  \includegraphics[width=0.7\textwidth]{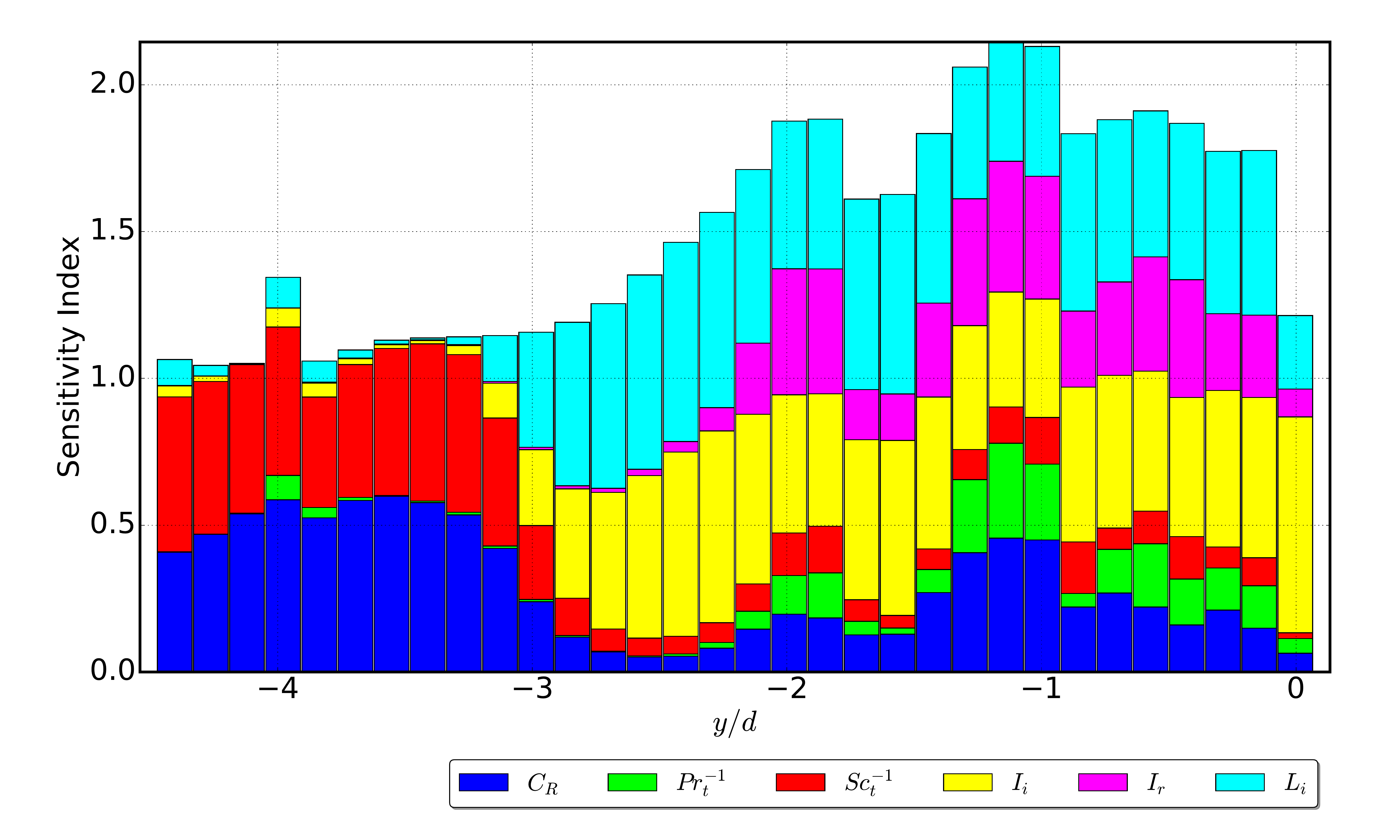}
  \caption{Total sensitivity indices for the calibration QOIs (scalar
    dissipation rate profile) for the 2D model. The more sensitive
    parameters are good candidates to embed the model error term.}
  \label{f:twoD_vs_threeD_sobol}
\end{figure}

\begin{figure}[htb]
  \centering
  \includegraphics[width=0.45\textwidth]{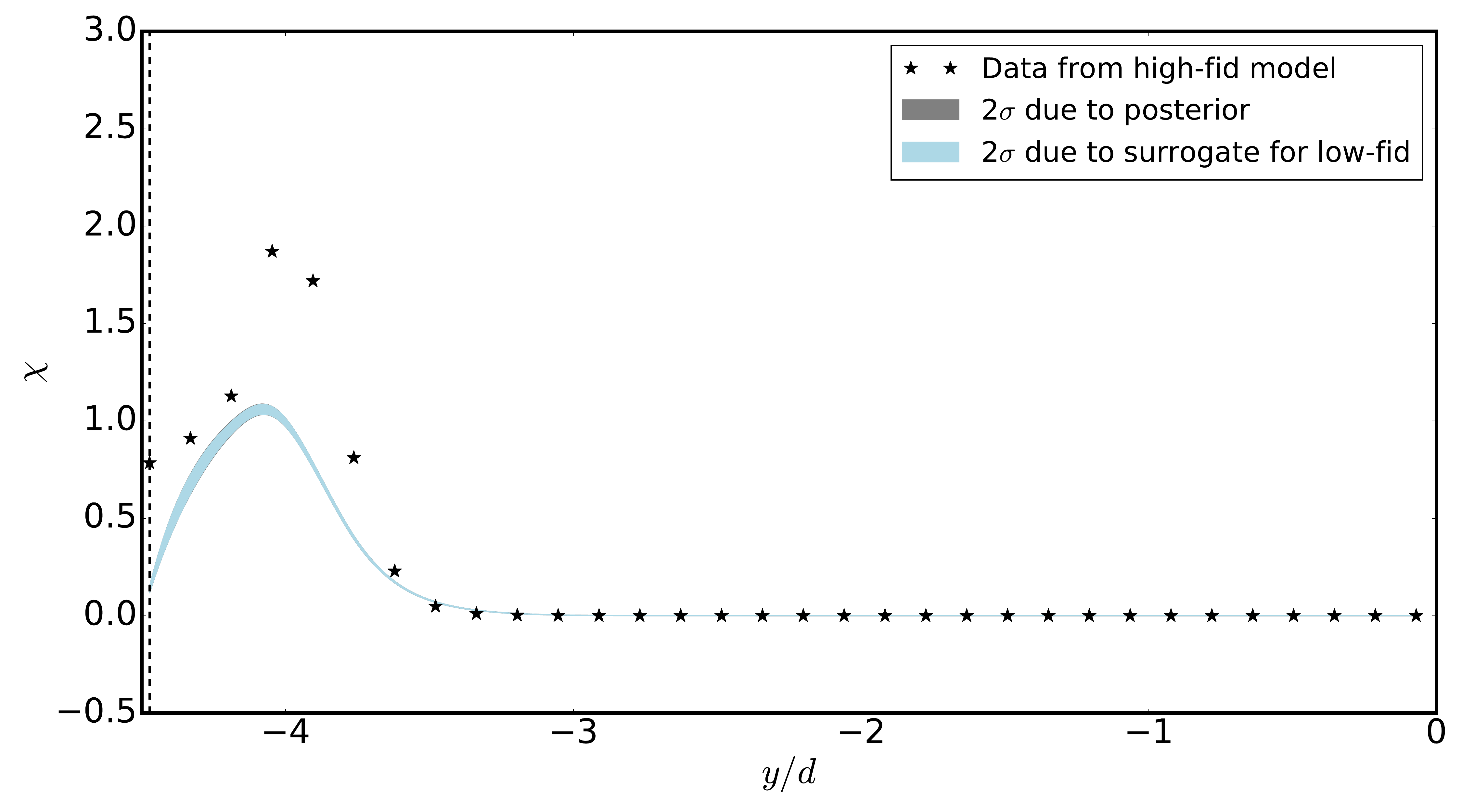}
  \includegraphics[width=0.45\textwidth]{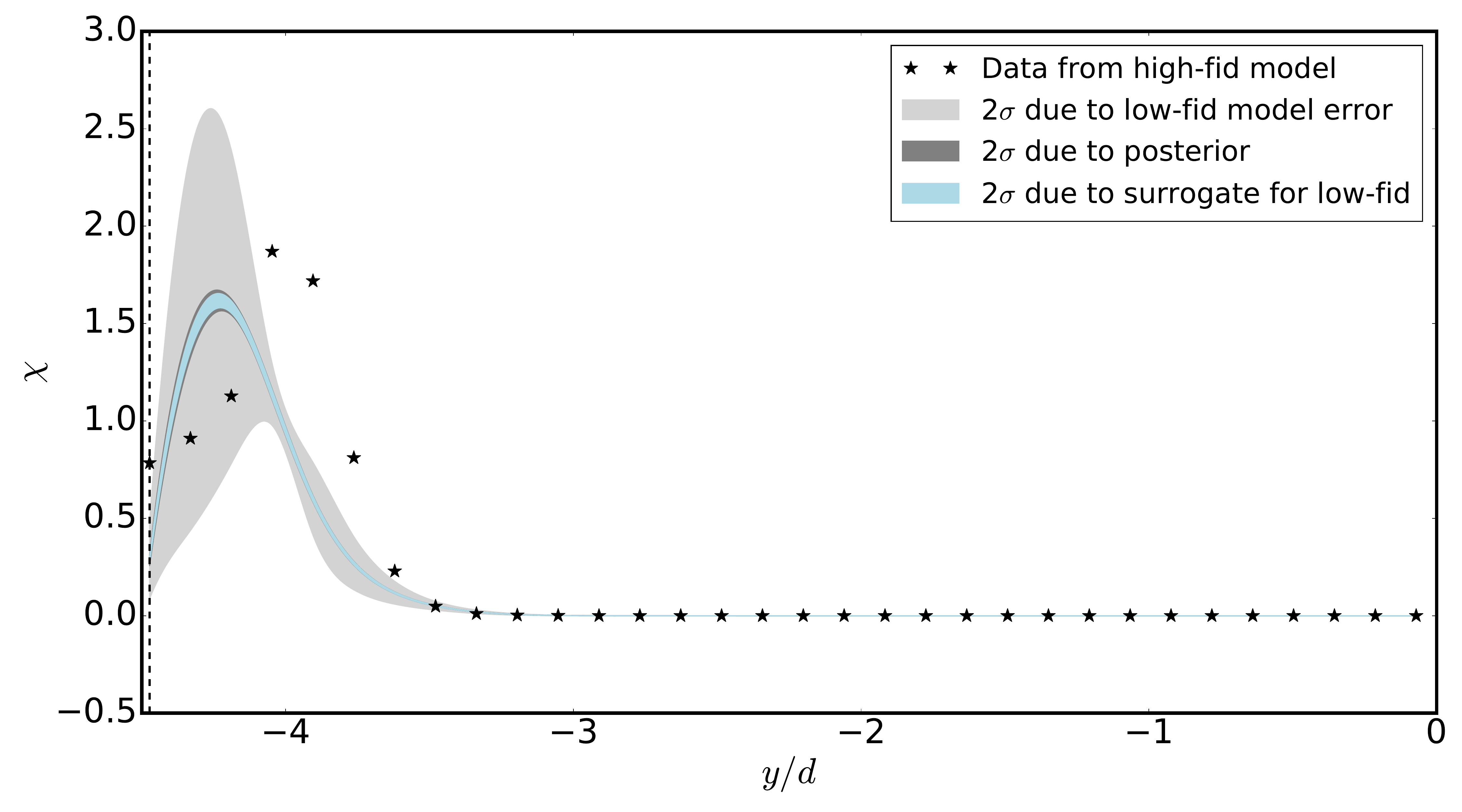}
  \includegraphics[width=0.45\textwidth]{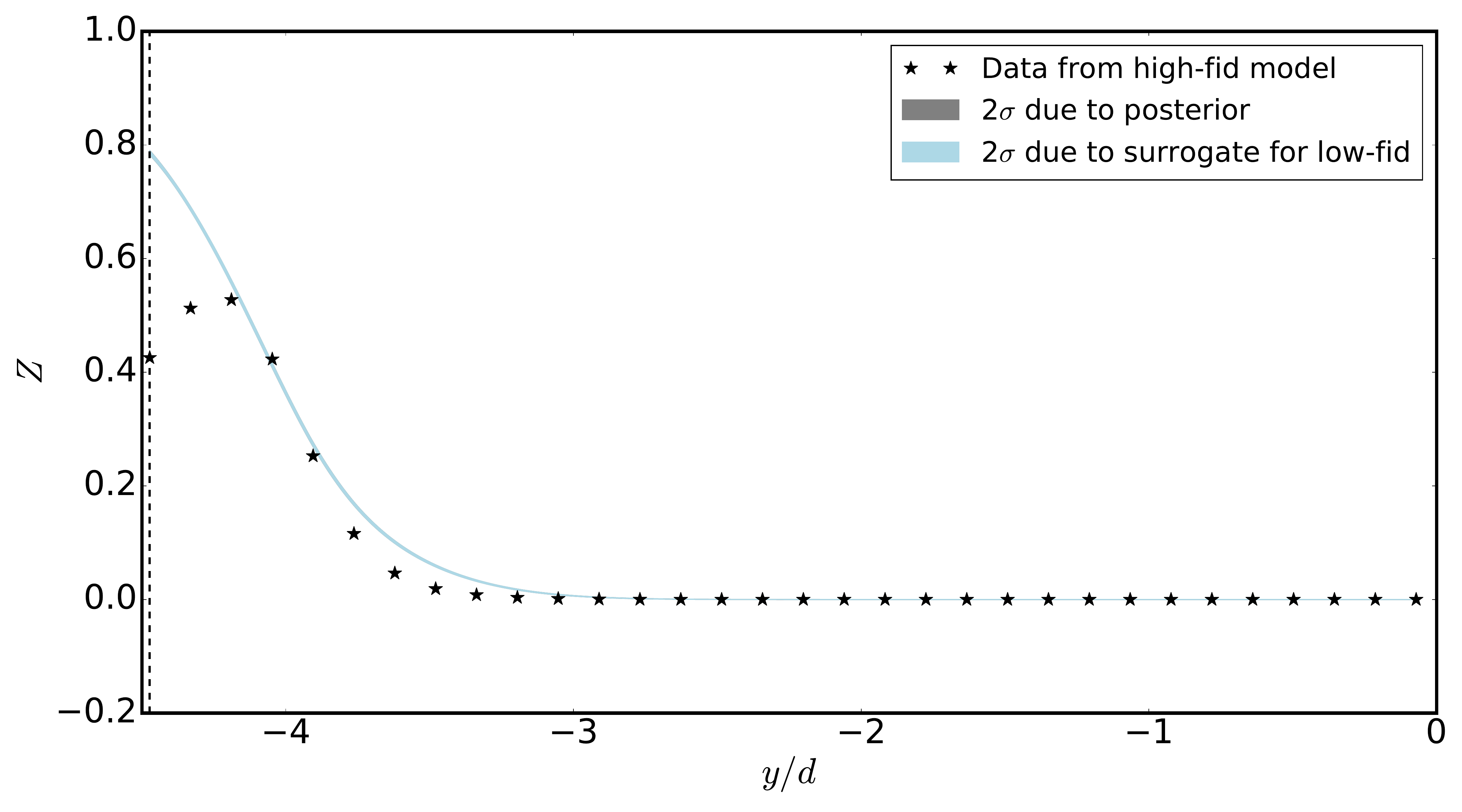}
  \includegraphics[width=0.45\textwidth]{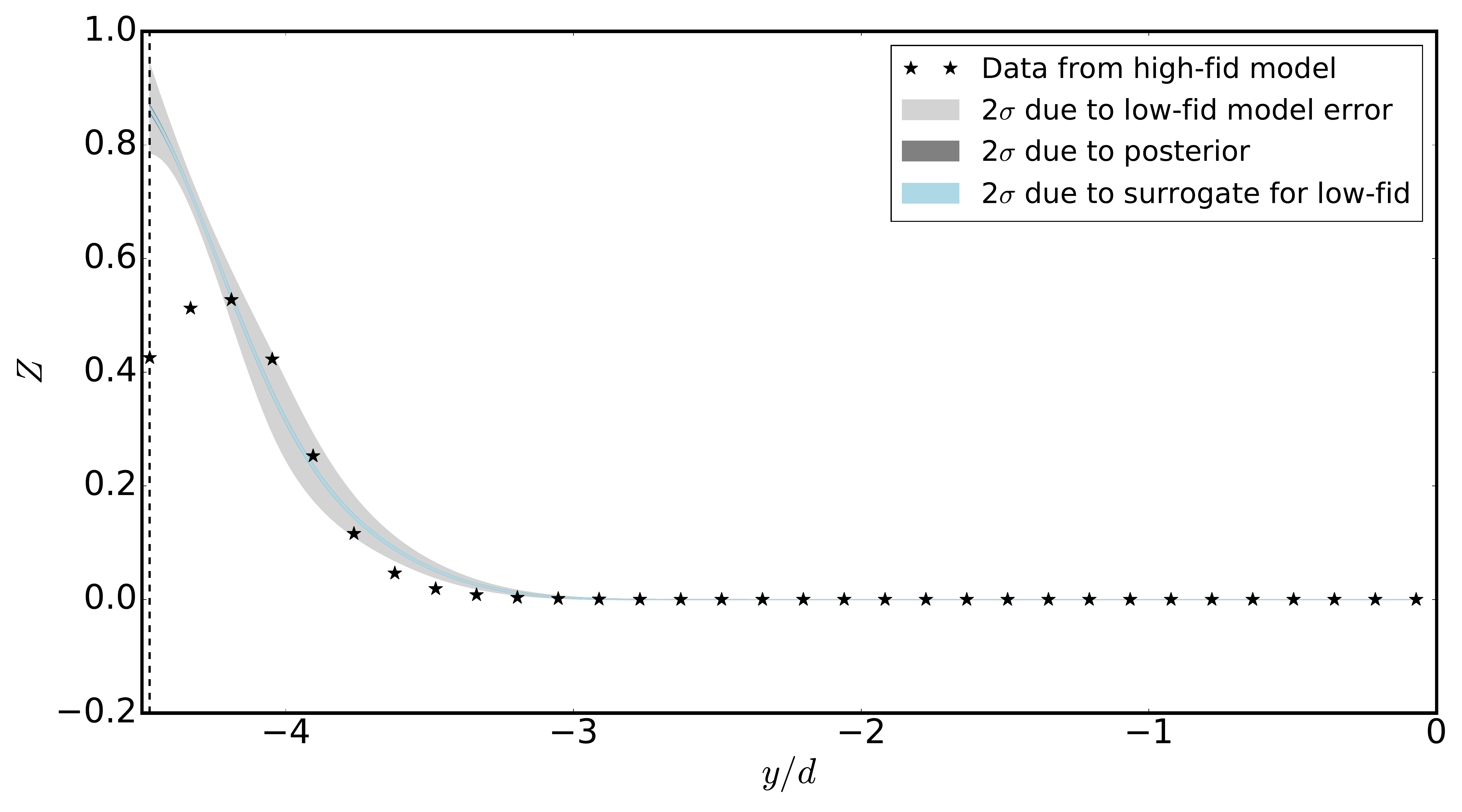}
  \includegraphics[width=0.45\textwidth]{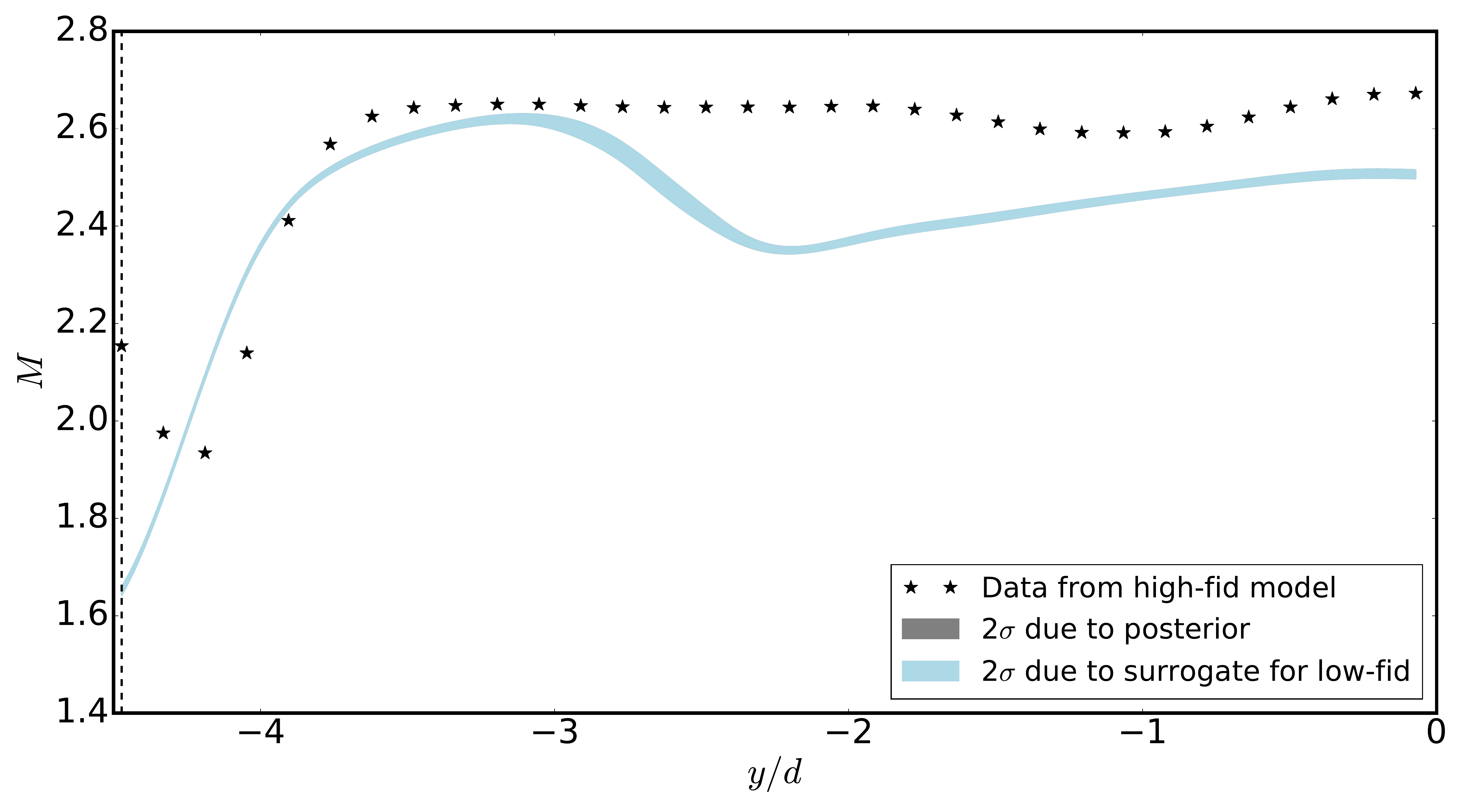}
  \includegraphics[width=0.45\textwidth]{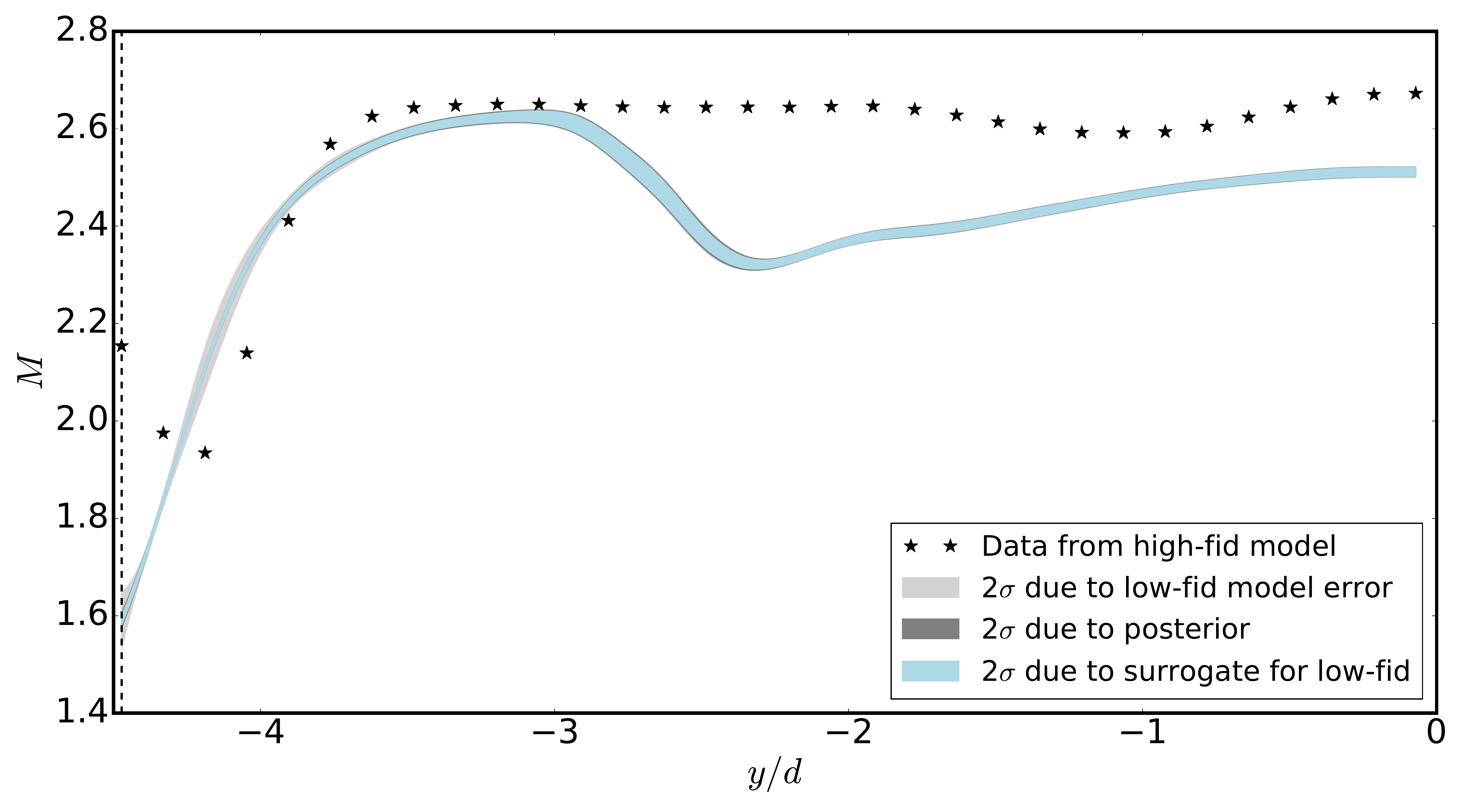}
  \caption{Posterior predictive distributions for $\chi$, $Z$, and $M$
    profiles from the 2D model without model error treatment (left
    column) and with embedded model error representation (right
    column).  }
  \label{f:twoD_vs_threeD_fullprofiles}
\end{figure}

We embed the model error representation in $C_R$ and $Sc_t^{-1}$,
allowing $\delta(\cdot)$ to be first-order Legendre-Uniform PCE for
these two parameters only, while all other parameters are treated in
the classical Bayesian inference sense (i.e., no embedding). A triangular structure
is enforced for the multivariate expansion of \cref{e:lambda_PCE},
which becomes
\begin{align}
  \[\begin{array}{l}C_R \\ Sc_t^{-1} \end{array}\]
  + \[\begin{array}{l} \alpha_{(1)} \xi_1 \\
  \alpha_{(1,0)} \xi_1 + \alpha_{(0,1)} \xi_2\end{array}\]
\end{align}
in accordance to the notation in \cref{e:PCEForm3}, and where we have
substituted the first-order Legendre-Uniform polynomial basis
$\psi(\xi)=\xi$. Priors with positive support are prescribed for
$\alpha_{(1,0)}$ and $\alpha_{(0,1)}$, which then guarantee a unique
distribution for each realization of the triple $(\alpha_{(1)},
\alpha_{(1,0)}, \alpha_{(0,1)})$.  The decision of where to embed is
guided by an initial GSA performed on the calibration QOIs: the $\chi$
profile. This is accomplished by using the low-fidelity surrogate
\cref{e:embed_with_surrogate} to estimate the total sensitivity
indices via the methodology in \cref{e:sobol_tot} for each of the
spatially-discretized $\chi$ grid points. Illustrated in
\cref{f:twoD_vs_threeD_sobol}, while sensitivity varies over $y/d$,
the overall most sensitive parameters, especially near the bottom wall
(left side of the plot) where $\chi$ values are far from zero (e.g.,
see \cref{f:twoD_vs_threeD_fullprofiles} first row), are $C_R$ and
$Sc_t^{-1}$. It is thus reasonable to expect the model error embedding
to be most effective when applied on these two parameters.  Indeed, in
separate studies (results omitted), embedding in other parameters
displayed less effective capturing of QOI discrepancy between the two
models.
The low-fidelity model surrogate is built using third-order Legendre
polynomial from 500 regression samples. The PCEs used for the
likelihood and posterior predictive are third-order Legendre-Uniform,
and integration over $\xi$ is performed using the 4-point Gaussian
quadrature rule in each dimension. MCMC is run for $10^5$ samples,
with 50\% burn-in and thinning of every 100 samples to improve mixing.

\begin{table}[htb]
\begin{center}
\begin{tabular}{cll}
\hline 
Parameter & Range & Description  \\\hline
$C_R$ & [0.005, 0.08]  & Modified Smagorinsky constant  \\
$Pr_t^{-1}$ & [0.25, 2.0] & Inverse turbulent Prandtl number \\
$Sc_t^{-1}$ & [0.25, 2.0] & Inverse turbulent Schmidt number \\
$I_{i}$ & [0.025, 0.075] & Inlet turbulence intensity magnitude in
horizontal direction  \\
$I_{r}$ & [0.5, 1.0] & Inlet turbulence intensity vertical to
horizontal ratio \\
$L_{i}$ & [0.0, 8.0] mm & Inlet turbulence length scale  \\
\hline 
\end{tabular}
\end{center}
\caption{Uncertain parameters for the 2D model in the model error
  study. }
\label{t:twoD_vs_threeD_params}
\end{table}

\Cref{f:twoD_vs_threeD_fullprofiles} depicts the 2D model posterior
predictive distributions for the profiles of scalar dissipation rate
$\chi$, mixture fraction $Z$, and Mach number $M$. The scalar
dissipation rate profile constitutes data used for calibration, while
other QOIs are extrapolatory. Overall, we see that the light grey band
covers the model-to-model discrepancy reasonably well for $\chi$, but
there are small regions where the high-fidelity data are uncovered
(e.g., around $y/d=-4$). For $Z$, we observe reasonable performance to
the right of the second grid point; and for $M$, the light grey band
is nearly non-existent (and thus has poor discrepancy coverage).

There are two important factors that explain these observed
limitations. The first reason is a challenge for all model error
approaches and not specific to the embedded representation: the
extrapolation of $\delta(\cdot)$ for use on QOIs outside the
calibration set (as described in \cref{e:embed_extrapolate}, e.g.,
extrapolated from $\chi$ to $Z$ and $M$ profiles in this example) may
be inadequate and lead to poor coverage of model-to-model discrepancy
from the uncertainty bands. This is one contributing factor for the
mismatched regions of the mid-right and bottom-right plots in
\cref{f:twoD_vs_threeD_fullprofiles}.
However, even in some regions of the calibration QOIs (top-right
plot), the light grey band does not extend to cover the high-fidelity
data points. This brings us to our second reason that also
demonstrates the limitation of the embedded representation: the model
error bands can only be as wide as the QOI ranges allowed by the
parameter variation (within the bounds of the uniform prior of
$\lambda$ in this case). This constraint presents a difficulty when
the QOI outputs from the low- and high-fidelity models are simply too
different, and cannot be compensated in the low-fidelity model by
varying its parameter values.
In our application, the 2D model is indeed physically very different
from the 3D in many aspects, and is unable to capture many detailed
physical features. (In contrast, the previous study of static and
dynamic Smagorinsky models presented much closer QOI behaviors.)  For
instance, a bow shock structure forms in the 3D setup, and would not
be portrayed in 2D. This difference can also change the locations
where the shocks reflect, thus yielding very different ``slice''
profiles. Furthermore, fuel injectors in 3D are circular (not
slotted), and not equivalent to a simple extrusion of the 2D
geometry. The shock strength is thus expected to be weaker in the 3D
model due to the relatively smaller area of fuel injection. These
insights are supported by the $M$ profile plots, where the shocks are
suggested by the dips in the profiles.
Additional studies (results omitted) indeed confirm that the posterior
predictive bands from the right column of
\cref{f:twoD_vs_threeD_fullprofiles} are similar to those of the prior
predictive, i.e., the widest allowable by the ranges in
\cref{t:twoD_vs_threeD_params}.  This lowered flexibility of capturing
model discrepancy is the price for requiring the governing equations
to be respected in the predictive results. At the same time, it also
inspires interesting future work directions. To begin with, we would
like to explore other forms of embedding that could be more
advantageous, both in terms of selection of $\lambda$ as well as
advanced structuring of $\delta(\cdot)$. Another possibility is to
construct a hierarchy of intermediate models that offers a smoother
transition between the high- and low-fidelity models, which also
complements the multi-model theme in \cref{ss:ml_mf}. We plan to
develop these techniques as we proceed to the full HDCR domain with
combustion enabled.

\section{Conclusions}
\label{s:conclusions}

The development of scramjet engines is an important area of research
for advancing hypersonic and orbital flights. Progress towards optimal
engine designs requires accurate flow simulations together with
uncertainty quantification (UQ). However, performing UQ for scramjet
simulations is extremely challenging due to the large number of
uncertain parameters involved and the high computational cost of
performing flow simulations. This paper addressed these difficulties
by developing practical UQ algorithms and computational methods, and
deploying them to a jet-in-crossflow problem in a simplified HIFiRE
Direct Connect Rig (HDCR) scramjet combustor.  To start, a
jet-in-crossflow test problem was formulated for the primary injector
section subdomain of the HDCR, with a focus on the interaction between
fuel jet and the supersonic crossflow without combustion. Large-eddy
simulation (LES) was used to model the turbulent flow physics, and the
fully coupled system of conservation laws was solved by the RAPTOR
code.

Global sensitivity analysis (GSA) was conducted to identify the most
influential uncertain input parameters, providing important
information to help reduce the system's stochastic dimension. GSA was
efficiently performed by leveraging multilevel and multifidelity
frameworks that combined evaluations from different models, polynomial
chaos expansion (PCE) surrogates that provided a convenient form for
calculating sensitivity indices, and compressed sensing that
discovered sparse PCE forms from limited simulation data. Through GSA,
six important input parameters were established from an initial set of
24.

A framework was then introduced for quantifying and propagating
uncertainty due to model error.  This technique involved embedding a
correction term directly in the parameters of the low-fidelity model,
thus guaranteeing the predictions to maintain satisfaction of the
underlying governing equations and physical laws. The correction term
was represented in a stochastic and Bayesian manner, and calibrated
using Markov chain Monte Carlo. Both the strengths and weaknesses of
this approach were highlighted via applications of
static-versus-dynamic Smagorinsky turbulent treatments as well as
2D-versus-3D geometries.

The logical next step is to extend these UQ techniques to the full
HDCR configuration (\cref{f:LES_full_schematic}). Additional
challenges are expected to emerge, both for LES involving a more
complex cavity geometry with combustion, and UQ that will face even
higher dimensional settings, increasingly expensive model evaluations,
and fewer data points. Additional numerical developments will be
essential to overcome these obstacles, with fruitful avenues of
exploration that include adaptive and robust quadrature methods,
Bayesian model selection, and efficient MCMC for model calibration.

\section*{Acknowledgments}

Support for this research was provided by the Defense Advanced
Research Projects Agency (DARPA) program on Enabling Quantification of
Uncertainty in Physical Systems (EQUiPS).
Sandia National Laboratories is a multimission laboratory managed and
operated by National Technology and Engineering Solutions of Sandia,
LLC., a wholly owned subsidiary of Honeywell International, Inc., for
the U.S. Department of Energy's National Nuclear Security
Administration under contract DE-NA-0003525.

\bibliography{local}
\bibliographystyle{siamplain}

\end{document}